# On the Axioms of Topological Electromagnetism

D.H. Delphenich[†]
Physics Department
University of Wisconsin – River Falls
River Falls, WI 54022

*Abstract. The axioms of topological electromagnetism are refined by the introduction of the de Rham homology of k-vector fields on an orientable manifold and the use of Poincaré duality in place of Hodge duality. The central problem of defining the spacetime constitutive law is elaborated upon in the linear and nonlinear cases. The manner by which the spacetime metric might follow from the constitutive law is examined in the linear case. The possibility that the intersection form of the spacetime manifold might play a role in defining a topological basis for the constitutive law is explored. The manner by which wave motion might follow from the electromagnetic structure is also discussed.*

**0. Introduction.** One of the long-standing problems of theoretical physics has been to discern whether the electromagnetic field can be obtained from the structure of the spacetime manifold, whether from its geometrical structure, like the gravitational field, or its topological structure, or some combination of both. Despite the many noble attempts at the unification of gravitation and electromagnetism (see Lichnerowicz [**1**] or Vizgin [**2**]), which usually took a purely geometric approach to the problem, only so much definitive progress was made. Einstein, in particular, felt that the unification of electromagnetism and gravitation would probably be closely related to two other long-standing problems: the causal interpretation of wave mechanics and the extension of electromagnetism to a nonlinear theory.

One of the consequences of the widespread interest in gauge field theories on the part of pure mathematics was the growing realization amongst physicists that topology also played an important role in all such theories, including what is perhaps the simplest gauge field theory of physics, namely, classical electromagnetism. Furthermore, if one returns to the purely physical considerations, one must remember that, historically, the introduction of the methods of differential geometry into physics, by way of general relativity, actually originated in the consideration of the structure of the characteristic submanifolds for the linear wave equation that followed as a special case of Maxwell's equations for electromagnetism. Consequently, there is something logically recursive about reintroducing the geometry of spacetime back into the foundations of electromagnetism.

A more logically straightforward approach would be to accept that the metric structure of spacetime is a *consequence* of the electromagnetic structure of spacetime and then try to be more precise about the nature of the phrase "electromagnetic structure." One might do well to accept the notion that the best structure to start with is a *G*-structure on spacetime [**3**], since the metric structure of spacetime is a particular example of such a structure. In particular, we shall treat electromagnetism as something that related to an *SL*(4)-structure on the spacetime manifold, i.e., a bundle of oriented unit-volume frames whose fundamental tensor field is the chosen volume element. However, the spacetime manifold is not assumed to be Lorentzian, *a priori*. For the sake of completeness, we include a brief synopsis of some relevant notions from the theory of *G*-structures.

---

[†] david.delphenich@uwrf.edu

Besides the axiomatization of electrodynamics, a second problem that one must confront is that of defining the reduction from the bundle of oriented unit-volume linear frames *SL*(*M*) to the bundle of oriented Lorentz frames *SO*(3,1)(*M*) in terms of consequences that follow from electromagnetic field theory. We mentioned that the spacetime metric should be the symbol of the wave operator that follows from the electrodynamics equations. It would also be sufficient to find a way of constructing a line field *L*(*M*) from the electrodynamical information. One should note that the motion of electromagnetic waves will not accomplish this, since the fact that they are, by definition, lightlike implies that no proper time parameterization or rest frame can exist for their motion. However, a wave-like solution to the Maxwell equations is represented by a 2-form *F* of rank 2. Consequently, such a 2-form defines a reduction *beyond* *SO*(3,1)(*M*) to an *SL*(2)-structure on *M*, i.e., an *SL*(2) reduction of the bundle of oriented unit-volume linear frames. This is because a 2-form *F* of rank 2 defines a two-dimensional subspace of $T_x(M)$ at each $x \in M$ in the form of the associated system to *F*, in the sense of Cartan [**4,5**].

The basic organization of this paper is as follows: First, we review some elementary notions from the theory of *G*-structures on the spacetime manifold. Then, we develop a homology theory for *k*-vector fields on orientable manifolds that is Poincaré dual to de Rham cohomology. Next, we briefly discuss how the de Rham Homology might relate to the concept of a current algebra. Finally, we introduce the axioms of topological electromagnetism and discuss various aspects of them, such as the role of gauge transformations, the deeper geometrical and topological nature of constitutive laws, and the existence of wavelike solutions.

**1. *G*-structures on spacetime.** If *G* is a subgroup of *GL*(*n*) and *M* is an *n*-dimensional differentiable manifold then a *G-structure* on *M*, which we denote by *G*(*M*), is a reduction of the bundle of linear frames on *M*, which we denote by *GL*(*M*), to a *G*-principal bundle. This means that each fiber $G_x(M)$ of *G*(*M*) can be obtained as the *G*-orbit of some frame $\mathbf{e}_x$ in the tangent space at $x \in M$. For example, the vierbein representation of a spacetime metric says that a metric at a point of spacetime is equivalent to the orbit of some linear frame under the action of the Lorentz group on *GL*(*M*). This example also illustrates the possible non-existence or non-uniqueness of the reduction.

If we regard the special Lorentz group *SO*(3, 1) as a subgroup of *GL*(4) then it might be more physically cautious and prudent to examine the consequences of reducing to this subgroup by way of a sequence of intermediate steps, such as:

(1.1) $\qquad GL(4) \leftarrow GL^+(4) \leftarrow SL(4) \leftarrow SO(3, 1)$.

At the group level, the first reduction amounts to choosing one of the two connected components to *GL*(4), namely, the component at the identity. The second reduction follows from choosing a scale reference for $\mathbb{R}^4$ in the form of a unit volume element. Topologically, the reduction from $GL^+(4)$ to *SL*(4) takes place by way of a deformation retraction. Hence, there is no change in the homotopy type of the Lie group under this reduction, which has the homotopy type of the maximal compact subgroup of $GL^+(4)$, namely, $SO(4)$. The third reduction then follows from the introduction of the Minkowski scalar product on $\mathbb{R}^4$, which changes the homotopy type to that of the maximal compact subgroup of $SO(3, 1)$, namely $SO(3)$.

At the level of reductions of the bundle of linear frames, the first reduction from $GL(M)$ to $GL^+(M)$ entails making a global choice of orientation for the linear frames on $M$. In the language of $G$-structures, an orientation on $T(M)$ can be regarded as a $GL(n)$-equivariant map from $GL(M)$ to $GL(M)/GL^+(M) = \mathbb{Z}_2$, which we represent b the set {+, −} with multiplication. In fact, in order to define an orientation, the map must be constant, since the bundle $GL^+(M)$ consists of all linear frames that map to the same element of $\mathbb{Z}_2$.

Of course, the existence of such a map depends upon the orientability of the tangent bundle, which points to the fact that reductions of $GL(M)$ may be obstructed by topological considerations. These topological considerations usually involve the Stiefel-Whitney classes of $T(M)$, which are certain $\mathbb{Z}_2$ cohomology classes for $M$. For compact manifolds, the obstruction to orientability amounts to the first Stiefel-Whitney class $w_1 \in H^1(M; \mathbb{Z}_2)$. In particular, if $M$ is simply connected then it is orientable; however, an elementary example in which the converse does not obtain is the torus. If $M$ is not orientable then there is a simply connected orientable covering manifold $\tilde{M}$ and a two-to-one covering map $\tilde{M} \to M$, which is a local diffeomorphism. Since a global section of this covering map is equivalent to an orientation, if $M$ is not orientable then any section will be undefined at some points of $M$. A minimal (in terms of inclusion) representation for such a set of singular points is given by the image of the singular $\mathbb{Z}_2$-3-cycle that is the Poincaré-Alexander dual of $w_1$, which represents a sort of "wall defect" in language of topological defects; we shall return to this notion shortly.

In general, a reduction of a $G$-structure to an $H$-structure is associated with either a section of the homogeneous space bundle $G/H(M) \to M$ or, equivalently, the inverse image of a coset of $G/H$ under a $G$-invariant map $G(M) \to G/H$. In the aforementioned case of an orientation, the bundle $G/H(M)$ is diffeomorphic to $\tilde{M}$. Often, as in the reduction of $GL(M)$ to a bundle of orthogonal frames, this latter map takes the form of a tensor field, such as the metric tensor field in the present example. The fundamental tensor field that is associated with the reduction from $GL^+(M)$ to $SL(M)$ is a global volume element for $M$, since the map $G(M) \to G/H$ takes the form of a global positive function on $GL^+(M)$:

(1.2)         det: $GL^+(M) \to GL+(n)/SL(n) = \mathbb{R}^+$,    $\mathbf{e} \mapsto \det(\mathbf{e})$

that implies the existence of a global non-zero $n$-form $\mathcal{V}$ on $M$:

(1.3)          $\mathcal{V}: M \to \Lambda^n(M)$,      $x \mapsto \det(\mathbf{e}_x)\theta^1 \wedge \ldots \wedge \theta^n$.

The coframe $\theta^i$, $i = 1, \ldots, n$ is the coframe that is reciprocal to the frame $\mathbf{e}_i$, so $\theta^i(\mathbf{e}_j) = \delta^i_j$. The order of the factors in the exterior multiplication is arbitrary, except insofar as it gives one or the other overall sign to the result, since the choice of sign was determined by the first reduction to $GL^+(M)$. One could also regard this fundamental tensor field as the "pseudoscalar" function that takes **e** to det **e**, although the prefix "pseudo" tends to hide the true nature of the difference between 0-forms and $n$-forms: they are equivalent only for frame transformations that do not change the sign of the determinant.

The actual reduction $SL(M)$ consists of all linear frames in $T(M)$ that map to +1 under det. Since the determinant of a frame represents the volume of the parallelepiped that is spanned by the frame, the bundle $SL(M)$ then consists of oriented linear frames that span a unit hyper-rectangular volume. Actually, once one has made the first reduction to $GL^+(M)$, there is no further topological obstruction to making the second one to $SL(M)$. This is because an obstruction would take its values in one of the homotopy groups of the homogeneous space $GL^+(M)/SL(M) = \mathbb{R}^+$, which is contractible.

The reduction from $SL(M)$ to $SO(3,1)(M)$ amounts to restricting oneself to oriented Lorentz orthogonal frames. Up to homotopy, this is equivalent to making a (non-canonical) choice of line field $L(M)$ on $M$. Mathematically, this is a particular type of line bundle that is obtained as a section of the projectivized tangent bundle to $M$. Physically, the line at any point represents the unoriented direction of proper time evolution at that point. Equivalently, if one restricts oneself to oriented Lorentz orthogonal frames that have one member that spans the line at each point, the $SO(3)$ orbit of those frames defines the *rest frame* at that point, relative to the motion that defined $L(M)$. For a non-compact manifold, there is no topological obstruction to this reduction, but for a compact manifold, the Euler-Poincaré characteristic of $M$, $\chi[M]$, must vanish. This, in turn can be related to the vanishing of the Euler class of $T(M)$, $e[M]$, or the top Stiefel-Whitney class $w_4$. Interestingly, that means the one-point compactification of Minkowski space, which is diffeomorphic to $S^4$, does not admit a Lorentzian metric.

Apparently, if we are to find a pre-metric formulation of electromagnetism in this sequence of reductions, it would have to involve the first three subgroups. The approach taken in this article will be to assume that we have the topological and geometrical machinery that one has accrued by the time that they reach $SL(M)$, namely, a choice of orientation and a global non-zero (unit) volume element. One might wish to consider the possibility that the source of the electromagnetic field is related to the topological obstructions to the construction of such a volume element, i.e., $w_1$. By Poincaré-Alexander duality, $w_1$ also corresponds to an element of the three-dimensional $\mathbb{Z}_2$ homology. A simple example of how $n-1$-dimensional $\mathbb{Z}_2$ cycles in compact $n$-dimensional manifolds can obstruct orientability is provided by the Möbius band, which produces an orientable submanifold when one removes a transverse line. From the standpoint of physics, it is intriguing to consider the source of an electromagnetic field to be a three-dimensional object, not a zero-dimensional one. Of course, in some cases, a three-dimensional object, such as an open ball, may be homotopically equivalent to a zero-dimensional one by contraction.

Conspicuous by its absence in the following discussion will be the introduction of a linear connection on $GL(M)$, the process of reducing it to an $SL(4)$ connection on $SL(M)$, or the physical nature of its curvature, torsion, and other geometric information. This may be related to the fact that by starting with more established axioms for electromagnetism, we are also introducing the Minkowski electromagnetic field strength 2-form $F$ with no further discussion of whether it actually has some more intrinsic geometrical construction. Of course, correcting that omission defines a very important direction for further research into the nature of electromagnetism.

**2. De Rham homology on orientable manifolds.** There are actually *two* ways by which the spacetime metric appears in the non-homogeneous Maxwell equations:

(2.1) $$dF = 0, \qquad d*F = -\frac{4\pi}{c_0}*J.$$

As is commonly recognized [**6**], the obvious one is by way of the Hodge *-operator. The second one is by the introduction of the "constant" $c_0$, which is actually derivable from deeper assumptions on the nature of the constitutive laws of the electromagnetic vacuum state. We shall discuss this latter aspect of Maxwell's equations in more detail later, but for the moment, we concentrate on the former one.

In effect, the Hodge * isomorphism follows from a more general isomorphism that one finds on orientable manifolds, in the form of Poincaré duality. The metric allows us to define a linear isomorphism between the tangent spaces to our manifold $M$ and the cotangent spaces, and in so doing, an isomorphism of the exterior algebra $\Lambda_*(M)$ of $k$-vector fields on $M$ with the exterior algebra $\Lambda^*(M)$ of exterior differential $k$-forms on $M$. In the absence of a metric, but the presence of a volume element $\mathcal{V}$, the only isomorphisms we can define are ([1]):

(2.2) $$*:\Lambda_k(M) \to \Lambda^{n-k}(M), \quad a \mapsto i_a \mathcal{V}.$$

In this expression, $a$ is a $k$-vector field on $M$, so it can be expressed as a finite sum of expressions of the form $X \wedge Y \wedge \ldots \wedge Z$, where $X, Y, \ldots, Z \in \mathcal{X}(M)$ are $k$ vector fields. The interior product of a $k$-vector field and an $p$-form $\alpha$ is defined in general by assuming that it is $k$-linear and recursively defining:

(2.3) $$i_{X \wedge Y \wedge \ldots \wedge Z}\, \alpha = i_X(i_{Y \wedge \ldots \wedge Z}\, \alpha).$$

As is well known, the exterior derivative operator $d$ on $\Lambda^*(M)$ makes $\Lambda^*(M)$ into a $\mathbb{Z}$-graded differential module (vector space, in fact). Since $d$ is of degree $+1$, one can define the *de Rham cohomology modules* by $H^k(M; \mathbb{R}) = Z^k(M; \mathbb{R}) / B^k(M; \mathbb{R})$, where $Z^k(M; \mathbb{R})$ consists of all of the closed $k$-forms and $B^k(M; \mathbb{R})$ consists of all exact $k$-forms. The exterior product of differential forms then gives rise to a ring structure on $H^*(M; \mathbb{R})$ whose product $[\alpha] \cup [\beta]$, which amounts to the usual "cup product," is derived from the exterior product:

(2.4) $$[\alpha] \cup [\beta] = [\alpha \wedge \beta].$$

As is less known, on an orientable manifold $M$ one can use the isomorphisms (2.2) and the operator $d$ to define a *codifferential* or "boundary" operator of degree $-1$ on $\Lambda_*(M)$ by way of:

---

[1] We temporarily revert to the general case of $n$-dimensional orientable differentiable manifolds.

(2.5) $$\delta: \Lambda_k(M) \to \Lambda_{k-1}(M), \quad a \mapsto \delta a = *^{-1}d*a.$$

A simple local computation shows that if $X \in \mathfrak{X}(M)$ then the components of $\delta X$ with respect to a natural local frame field $\mathbf{e}_i = \partial/\partial x_i$ are going to agree with the usual divergence of $X$ that one learns from vector calculus. An important point to emphasize is that, unlike the usual definition of the divergence operator in differential geometry, we have not needed to introduce a metric for ours.

It is not hard to see that the fact that $d^2 = 0$ implies that:
(2.6) $$\delta^2 = 0.$$

Hence, $\delta$ makes $\Lambda_*(M)$ into a $\mathbb{Z}$-graded differential module. One then defines a homology by way of:
(2.7) $$H_k(M; \mathbb{R}) = Z_k(M; \mathbb{R}) / B_k(M; \mathbb{R}),$$

in which $Z_k(M; \mathbb{R})$ is the space of a *co-closed* $k$-vector fields ($\delta a = 0$) and $B_k(M; \mathbb{R})$ consists of all *co-exact* $k$-vector fields ($a = \delta b$ for some $k+1$-vector field $b$).

A significant difference between the behavior of $H_*(M; \mathbb{R})$ and $H^*(M; \mathbb{R})$, besides the degree of the boundary map, is that the exterior product of a $p$-vector field $a$ and a $q$-vector field $b$ does not "descend to homology" since $\delta(a \wedge b)$ does not generally equal $\delta a \wedge b + (-1)^p a \wedge \delta b$. One does have the following useful relation for the product of a smooth function $f$ and a $k$-vector field $a$:
(2.8) $$\delta(fa) = *^{-1}(df \wedge *a) + f \delta a.$$

When $a$ is a vector field, this becomes:
(2.9) $$\delta(fa) = af + f \delta a.$$

Before one suspects that something is missing in our homology, keep in mind that one does not usually expect to find a ring structure on homology, at least in general. Nevertheless, one does find that the interior product of $k$-vector fields and $p$-forms ($p > k$) descends to (co)homology, where it takes the form of the usual "cap product" that one encounters in the topology of manifolds [**7-10**]:
(2.10) $$H_k(M; \mathbb{R}) \times H^p(M; \mathbb{R}) \to H^{p-k}(M; \mathbb{R}), \quad ([a], [\beta]) \mapsto [a] \cap [\beta] = [i_a \beta].$$

In this definition, the [ ] brackets denote the homology or cohomology class that corresponds to the item inside.

When one descends to homology the cap product produces the isomorphisms of *Poincaré duality:*
(2.11) $$H_k(M; \mathbb{R}) \cong H^{n-k}(M; \mathbb{R}), \quad [a] \mapsto [a] \cap [^0V] = [*a].$$

Since our coefficient ring is a field, it is correct to identify $H^k(M; \mathbb{R})$ as the dual of the vector space $H_k(M; \mathbb{R})$; i.e., $H^k(M; \mathbb{R}) \cong \mathrm{Hom}(H_k(M; \mathbb{R}); \mathbb{R}) = H_k(M; \mathbb{R})^*$. Hence, its elements can be regarded as linear functionals on homology classes. When M is compact, or we restrict ourselves to *k*-vector fields of compact support, we can represent the *k*-dimensional cohomology class [α] as a linear functional on *k*-dimensional homology classes [*b*] in integral form:

(2.12) $$\alpha[b] = \int_M \alpha \wedge *b = \int_{\tilde{b}} \alpha.$$

However, in order for the last integral to make sense, we use of the fact that de Rham's theorem for cohomology, which we consider to be an isomorphism of the de Rham cohomology of *M* with the singular cubic cohomology with real coefficients, also gives rise to a corresponding isomorphism of de Rham homology with singular cubic homology with real coefficients. Hence, the homology class [*b*] also corresponds to a *k*-dimensional singular cubic homology class [$\tilde{b}$] – which is represented by some closed singular cubic *k*-chain $\tilde{b}$ − over which we can integrate. If the support of the *k*-vector field *b* is the image of the *k*-chain $\tilde{b}$ then one can think of *b* as a *k*-vector field on $\tilde{b}$.

This seems to be of immediate relevance to the investigation of the deeper nature of the source currents of physical fields. For instance, one associates real numbers (charges) with point sources and vector fields (currents) with line sources. Of course, one also associates real numbers with higher-dimensional *k*-chains, such as charge densities, but that might also relate to their contractibility. For instance, stable currents only flow in conducting loops, which are closed 1-chains.

We also have the homological form of the *Poincaré Lemma:* every point of *M* has a neighborhood on which any co-closed *k*-vector field is co-exact. Indeed, one need only find the image of an open ball about the point in some coordinate chart.

In order to facilitate the physical interpretation, we refer to the homology class of a vector field *J* as a *conserved current* and the corresponding cohomology class in dimension *n*−1, [*\*J*] = [$i_J \mathcal{V}$], as its *flux density*. For any *n*−1-dimensional submanifold *S*, the integral:

(2.13) $$J[S] = \int_S *J$$

is the *total flux of J through S*. This allows us to give an integral form to our differential requirements on *J*, namely:

(2.14a)      δ*J* = 0       iff       *J*[*S*] = 0 for all *S* such that *S* = ∂*V*
(2.14b)      *J* = δ*B*     iff       *J*[*S*] = 0 for all *S* such that ∂*S* = 0.

This can also be accounted for using Stokes's theorem (more precisely, *Gauss's theorem*) in the following form:

(2.15)                         δ*J*[*V*] = *J*[∂*S*].

One should observe that although we have sacrificed the ring structure of cohomology for the more limited structure of homology, nevertheless, in the present case, since our homology classes are represented by vector fields we also inherit the structure of a Lie algebra on the one-dimensional homology classes. This is because, as is straightforward to verify, the Lie bracket of conserved vector fields is conserved:

(2.16)              if δ*X* = 0 and δ*Y* = 0 then δ[*X*, *Y*] = 0.

This is essentially the statement that the infinitesimal generators of volume-preserving diffeomorphisms on an orientable manifold define a Lie algebra. Furthermore, one can define an action of the conserved currents on the homology classes and cohomology classes by way of the Lie derivative operator:

(2.17a) $\qquad L_{[X]}[a \wedge b \wedge \ldots \wedge c] = [L_X (a \wedge b \wedge \ldots \wedge c)]$

where:

$$L_X (a \wedge b \wedge \ldots \wedge c) = [X,a] \wedge b \wedge \ldots \wedge c + a \wedge [X,b] \wedge \ldots \wedge c + a \wedge b \wedge \ldots \wedge [X,c]$$

(2.17b) $\qquad L_{[X]}[\alpha] = [L_X \alpha]$

where:

$$L_X \alpha = d i_X \alpha + i_X d\alpha.$$

**3. Current algebras.** Since the currents that we defined in the previous section are closed under Lie bracket, we are naturally tempted to call that Lie algebra the "current algebra" for an orientable $M$. However, the term "current algebra" is already well-established in strong interaction physics, where it refers to a Lie algebra of field operators that also contain quantum information in their algebraic structure, and has also more recently [**11**] been associated with Lie algebras of smooth maps $M \to \mathfrak{g}$ from a manifold $M$ into a Lie algebra $\mathfrak{g}$ that represents the infinitesimal generators of internal − i.e., gauge − symmetries of the fields that one is concerned with.

This latter conception of current algebras admits an immediate representation when $\mathfrak{g}$ is the Lie algebra of a Lie $G$ that serves as the structure group of a $G$-principal bundle $P \to M$. One simply maps each $a(x) \in \mathfrak{g}$ for $x \in M$ to the fundamental vector field $\tilde{a}_x$ on $P_x$, which is $G$-invariant and vertical. These vector fields generate a group of bundle automorphisms of $P$ that is generally referred to as the group of *gauge transformations*, and can also be represented by the group of smooth maps $M \to G$.

In order to reconcile our currents with these currents, we need to consider the Lie algebra of vertical $GL(n)$-invariant vector fields on the bundle $GL(M)$ of linear frames on $M$ that infinitesimally preserve a volume element $\tilde{\mathcal{V}}$ on $GL(M)$; i.e. $X \in \mathfrak{X}(GL(M))$ such that:

(3.1) $\qquad L_X \tilde{\mathcal{V}} = d i_X \tilde{\mathcal{V}} = 0,$

which is, of course, equivalent to:

(3.2) $\qquad \delta X = 0.$

The most natural volume element to define on $GL(M)$ is:

(3.3) $\qquad \tilde{\mathcal{V}} = \theta^1 \wedge \ldots \wedge \theta^n \wedge \omega^1 \wedge \ldots \wedge \omega^{n^2},$

in which $\theta^i$ is the canonical 1-form on $GL(M)$ and $\omega$ is the 1-form of a linear connection on $GL(M)$, where we have indexed the matrix elements of $\omega$ in column-major order, for the sake of specificity. This Lie algebra can then be represented more concisely by smooth maps from $M$ to $\mathfrak{gl}(n)$.

If one goes this route with current algebras, one notes that there is a snag associated with defining energy-momentum as an incompressible vector field on *GL*(*M*). That current is usually associated with symmetries under the translation group of $\mathbb{R}^n$, which only acts vertically on the fibers of the *affine* frame bundle over *M*. If one wishes to represent $\mathbb{R}^n$ on *GL*(*M*) by horizontal vector fields, such as the basic vector fields $\tilde{E}_i$ that are reciprocal to $\theta^i$ by way of $\omega$, then one must also note that if the connection $\omega$ is not flat then this representation is not faithful, or even a true representation, since the commutation rules of the $\tilde{E}_i$ are those of the infinitesimal generators of the Lie *pseudo*group of (germs of) local *parallel* translations of the frames on *M*:

(3.4) $$[\tilde{E}_i, \tilde{E}_j] = -\Theta^k(\tilde{E}_i, \tilde{E}_j)\tilde{E}_k - \Omega^l_m(\tilde{E}_i, \tilde{E}_j)\tilde{E}^m_l.$$

In this expression we have introduced the fundamental vertical vector fields $\tilde{E}^m_l$ on *GL*(*M*) that are reciprocal to the 1-forms $\omega^m_l$. We can see that if $\omega$ has zero curvature then its torsion represents a deformation of the Abelian Lie algebra of $\mathbb{R}^n$ into a non-Abelian Lie algebra, and if $\omega$ has zero torsion then its curvature represents a deformation of the finite-dimensional Abelian Lie algebra $\mathbb{R}^n$ into a larger non-Abelian subalgebra of $\mathfrak{X}(GL(M))$ that includes vertical vector fields and is generated by the $\tilde{E}_i$, which no longer define a basis, though.

Let us express our volume element on *GL*(*M*) in the form:

(3.5) $$\tilde{\mathcal{V}} = \tilde{\mathcal{V}}_H \wedge \tilde{\mathcal{V}}_V,$$

where:

(3.6) $$\tilde{\mathcal{V}}_H = \theta^1 \wedge \ldots \wedge \theta^n,$$

represents a volume element on the horizontal sub-bundle *H*(*GL*(*M*)) of *T*(*GL*(*M*)) that $\omega$ defines and:

(3.7) $$\tilde{\mathcal{V}}_V = \omega^1 \wedge \ldots \wedge \omega^{n^2},$$

defines a volume element on the vertical sub-bundle *H*(*GL*(*M*)).

In this form we can see that when *X* is a horizontal vector field in order for it to infinitesimally preserve $\tilde{\mathcal{V}}$ it is necessary and sufficient that:

(3.8) $$L_X \tilde{\mathcal{V}}_H = 0.$$

In order for this to be true, it is sufficient that:

(3.9) $$0 = L_X \theta^i = i_X d\theta^i + di_X \theta^i = i_X \Theta^i - \omega^i_j X^j + dX^i.$$

Such an *X* represents an infinitesimal parallel translation of the coframe $\theta^i$. Note that the basic vector fields fall into this category.

When *X* is vertical, in order for it to infinitesimally preserve $\tilde{\mathcal{V}}$ it is necessary and sufficient that:

(3.10) $$L_X \tilde{\mathcal{V}}_V = 0.$$

A sufficient condition for this is that:

(3.11) $\qquad\qquad\qquad 0 = L_X \omega = i_X d\omega + di_X \omega = i_X \Omega - i_X (\omega \wedge \omega) + di_X \omega.$

Such an $X$ is also called an *infinitesimal affine transformation* ([2]).

**4. Axiomatic topological electromagnetism [12-14].** We assume that we are given an orientable four-dimensional differentiable manifold $M$ with a global volume element $\mathcal{V}$. The basic data of electromagnetism consists of a smooth vector field $J$, which we call the *source current*, and a 2-form $F$, which we call the *electromagnetic field strength 2-form*.

Because we are trying to avoid all metric-related assumptions, we note that since we cannot distinguish statics from dynamics without the imposition of a rest frame for the motion in question, we must deal with field sources in the most general case, for which the field source is a conserved electric current.

One might easily argue that the vector field $J$ itself defines a line field over its own support supp($J$), which represents a partial Lorentzian structure. Physically, it represents the equivalence class of Lorentz frames at each point of supp($J$) in which the components of $J$ with respect to all but one leg are zero. In such a frame, $J$ can be the source of only electrostatic fields, since it has zero spatial velocity in its own rest frame. The only way that a current $J$ can feel a magnetic field in its own rest space is if $J$ is not the source of that field. This fact will prove crucial in our discussion of the Lorentz force.

Note that any conserved electric current $J$ can be decomposed into a product:

(4.1) $\qquad\qquad\qquad J = \rho \mathbf{v}$

in which $\rho$ is a smooth non-negative function that represents a charge density and $\mathbf{v}$ is a smooth vector field that represents the velocity flow field of the charge-bearing matter, although not uniquely. The observation is really more in the nature of a physical interpretation of the conserved electric currents as being reducible to kinematical entities. Note that $\mathbf{v}$ can still be compressible, even though, by definition, $J$ is not. This decomposition introduces a subtle physical equivalence in the form of the set of all ($\rho$, $\mathbf{v}$) such that $J = \rho \mathbf{v}$ for a given incompressible vector field $J$. This set, in turn, is clearly parameterized by the set of all smooth positive functions on the support of $J$, which also forms a group under multiplication. We shall return to this group later in the context of dilatational gauge symmetries.

One should observe at this point that – at least in the eyes of electromagnetism – it is the *current* that is fundamental ([3]), not the decomposition into $\rho$ and $\mathbf{v}$.

Since we cannot demand that $J$ have compact spacelike support, we might also consider a class of $J$ whose support is on "thickened curves." These would be 4-chains that have 1-chains as deformation retracts, such as four-dimensional cylinders of the form $(0, 1) \times B^3$, or solid torii of the form $S^1 \times B^3$, where $B^3$ is an open 3-ball. Such a $J$ could be called an *elementary current.* In the former case, one could go one step further and contract the curve to a point; in the latter case, whether one could contract $S^1$ to a point would depend upon whether $M$ was assumed to be simply connected. In terms of homology, we are asking the question of whether a given 1-chain is a 1-cycle, and, if so, does it bound. This suggests a way of distinguishing static from dynamic currents, at least in the eyes of topology. Since we are dealing with differentiable 1-chains, we are also allowed the luxury of a naturally defined vector field along them, which could play the role of **v** above. Note that this also forces us to consider only $J$'s that are obtained from **v** by a conformal factor.

---

[2] Strictly speaking, this term refers to a vector field on $M$ whose *lift* to $GL(M)$ has that property, but we shall take the viewpoint that it is more convenient to define all of our geometric objects on $GL(M)$ directly.

[3] Once again, we are making a thinly veiled attempt to endow our theory with a learned borrowing from the theory of current algebras in quantum chromodynamics.

As for the support of $F$, we make no assumptions at this point. However, one should keep in mind some of the considerations of the classical theory of electromagnetism, such as the way that the field strengths are not actually defined at source points, although they can be defined inside a spatially extended charge or current distribution. Similarly, causality considerations generally dictate that electromagnetic waves are only defined on light cones. However, we are trying to adhere to a strict regimen of relegating all metric considerations to a corollary status.

Ultimately, a complete theory of electromagnetism should describe the nature of the fields $J$ and $F$ and the way that they are coupled to each other, and possibly themselves. For instance, we need to know how the source current $J$ produces the field $F$. Generally, this relationship takes the form of a partial differential equation, such as Poisson's equation. It can also take the form of an integral equation, as with Huygens's principle, which one can obtain as a fundamental solution to the differential equation. Of particular interest are the "radiative" solutions, which are presumably due to the possibility that $J$ might be time varying, such as when **v** is also time varying. A key point to resolve is whether such radiative solutions are produced by a $J$ that is associated with a constant acceleration in **v**. If they are not, then we have another internal symmetry to the field theory, which relates to conformal symmetries that are more involved than homotheties, namely, the nonlinear inversion symmetries that correspond to the transformations from an inertial frame to a frame that moves with constant relative acceleration. Once again, we point out that if $J$ is the source of $F$ then it is entirely possible that a time variation in ρ can cancel a time variation in **v** in such a way that $J$ has no time variation and produces no radiative $F$ as a consequence. (However, since this only pertains to time variations within the line spanned by **v**, this does not seem to suggest any immediate application to the problem of canceling the centripetal acceleration of atomic electrons.)

Conversely, if another electric current $J'$ is in the field $F$ that is produced by a source $J$, one expects that there should be a coupling of $F$ to $J'$. In its simplest form, this is the Lorentz force law. However, the fact that accelerated charges might produce radiation in reaction to this force – i.e., a field $F'$ – suggests that the situation is more involved, especially when one considers a nonlinear superposition of $F'$ with $F$.

There is also the question of how the field $F$ that is produced by one current $J$ interacts with the field $F'$ of another current $J'$. Whether or not the interaction between the fields $F$ and $F'$ is one of simple superposition (tensor addition or what have you) depends on the degree of linearity in the field equation that couples currents to fields, since we could also say that the resulting field must be a solution of the same field equation when one uses the combination of *J* and *J′* as source. We shall not elevate this statement to the status of an axiom since it represents simply a physical interpretation of the solution to the field equations when the source is *J* + *J′*. In the following set of axioms, the degree of linearity in the field equation depends upon the nature of the constitutive law that one assumes for the medium in which the fields are defined.

Finally, if we are to leave open the possibility of nonlinear electromagnetism, we should seriously consider the possibility of self-interactions for *J* and *F*. The problem of defining a stable static charge distribution that is not pointlike seems to demand the introduction of a self-interaction of the charge distribution that is attractive at short distances and dies away much faster than $1/r^2$. Similarly, the fact that an accelerating charge – even in the absence of an external field – will presumably radiate energy, which amounts to a decelerating force that acts on the charge, also represents a type of indirect self-interaction of *J* itself by way the intermediary of *J* producing a radiation field which then interacts with *J* as a Lorentz-type force.

The interaction of *F* with itself might take the form of a "saturation limit" term for the field strengths or something more elaborate that produces the phase transition of vacuum polarization past that limit. In fact, the usual statement of the Klein paradox amounts to asking how one is supposed to deal with self-interactions of the electromagnetic field under vacuum polarization without simply expanding one's scope to a nonlinear theory. The notion that there should be a saturation limit for electric field strengths as a result of vacuum polarization is at the basis for the Born-Infeld model [**15**] for nonlinear electrodynamics.

*Axiom* 1. *Conservation of charge:* *J* defines a homology class in $H_1(M; \mathbb{R})$, i.e.:
(4.2) $\qquad \delta J = 0.$

*Axiom* 2. *Conservation of magnetic flux:* *F* defines a cohomology class in $H^2(M; \mathbb{R})$, i.e.:
(4.3) $\qquad dF = 0.$

*Axiom* 3. *Constitutive law:* There exists a smooth map that takes one of the following forms:
(4.4a) $\qquad f: \Lambda^2(M; \mathbb{R}) \to \Lambda_1(M; \mathbb{R}), \qquad F \mapsto J$
(4.4b) $\qquad f: \Lambda_1(M; \mathbb{R}) \times \Lambda^2(M; \mathbb{R}) \to C^\infty(M), \qquad f(F, J) = 0.$

*Physical interpretation* 1. *Lorentz force:* The 1-form $i_J F$ represents the force that is exerted by the electromagnetic field *F* on the current *J*. When *J* is the source of *F* this force is that of radiative reaction.

*Physical interpretation* 2. *Superposition:* When *F* is the field produced by a source *J* and *F′* is the field of a source *J′* then the field that is produced by the source *J* + *J′* ([4]) represents the (possibly nonlinear) superposition of the fields *F* and *F′*.

*Physical interpretation* 3. *Constitutive law.* If the operator $f = \delta\chi$ for some diffeomorphism $\chi: \Lambda^2(M; \mathbb{R}) \to \Lambda_2(M; \mathbb{R})$ then the equations:

(4.5)  $\mathfrak{h} = \chi(F)$ and $\delta\mathfrak{h} = J$

complete the usual Maxwell equations, in which the nonlinearity originates in the constitutive law $\chi$ and reduces to linearity in the limiting case of weak field strengths.

In the paper of Hehl, Obukhov, and Rubilar [**12**], two other axioms were mentioned. In addition to the aforementioned three axioms, it was suggested that one would need to postulate the existence of a proper time simultaneity foliation [**16**] and raise the existence of a Lorentz force from the status of a physical interpretation to a mathematical axiom. In the present formulation, the latter axiom is unnecessary because if *J′* is another conserved current that is in the field *F* produced by the source current *J* then we can form the interior product:

(4.6)  $\alpha = i_{J'} F$

with no further assumptions.

As for the axiom concerning the existence of a proper time simultaneity foliation, one should point out that such a construction is usually an artifact of defining a Lorentzian structure, by way of a rest frame. Since we are trying to exhibit the reduction from the *SL*(4) structure that facilitates electromagnetism to the bundle of oriented Lorentz frames as a consequence of the axioms of electromagnetism, we shall try to avoid introducing the axiom of such a foliation from the outset.

However, we can say this much about Lorentzian structures and rest frames, *a priori:* All it takes to define a Lorentzian structure, at least in the eyes of homotopy theory, is a line field, and one does have such a Lorentzian structure on the set of points, S(*J*), at which *J* is non-zero; since this set is an open subset of *M*, it is also a manifold. The line $L_x(S)$ at every point $x \in S(J)$ is then the one that is spanned by *J* at that point. Furthermore, because the line field $L_x(S)$ was defined by a non-zero vector field to begin with, it is an orientable line bundle. As for the matter of defining a Lorentzian pseudo-metric, it is important to note that the homotopy equivalence of a global line field on *M* with a Lorentzian pseudo-metric *g* does not define a *canonical* association of one with the other. In effect, any 3-plane in each tangent space to $x \in S(J)$ that does not contain the line $L_x(S)$ could serve as an orthogonal complement.

The line bundle $L(S) \to S(J)$ also defines a rest frame for the motion of the charge distribution that defines the current, namely the equivalence class of unit-volume linear frames at each point of S(*J*) that have one leg in $L_x$(S). Since we have not specified which Lorentzian metric is associated with *L*(S) we can only define a reduction of *SL*(*M*) to *SL*(3)(*M*) as a result of this. Similarly, without a choice of *g* we have no unique choice of complementary sub-bundle to *L*(S) in *T*(S). Hence, the integrability of the 3+1 decomposition of *T*(S) into a foliation of S by proper time simultaneity leaves becomes moot.

---

[4] The sum in this expression refers to the addition of 1-chains. Of course, the carrier of the sum is the union of the carriers and the coefficients at the points of their intersection is the vector sum, so this is still a vector addition, after all.

Notice that when we state the integral forms of the first two axioms, namely:

(4.7a) $\quad J[V] = \int_V *J = 0 \quad$ for any bounding three-dimensional region *V*

(4.7b) $\quad F[S] = \int_S F = 0 \quad$ for any bounding two-dimensional region *S*,

we cannot make the usual stipulations about whether these regions are spacelike, since we are trying to relegate such purely Lorentzian concerns to *consequences* of the electromagnetic structure.

If we look at the first axiom in differential form, we observe that when $J = \rho \mathbf{v}$ it also takes the form of a continuity equation:

(4.8) $\quad\quad\quad\quad 0 = \delta J = d\rho(\mathbf{v}) + \rho \delta \mathbf{v}$

or:

(4.9) $\quad\quad\quad\quad \mathbf{v}\rho = -\rho\delta\mathbf{v},$

which can also take form:

(4.10) $\quad\quad\quad\quad \delta\mathbf{v} = -\mathbf{v}(\ln \rho).$

This shows that **v** is incompressible iff ρ is constant along the flow of **v**. We shall return to the formal structure of equation (4.10) in a different context.

The integral form of the first axiom then becomes:

(4.11) $\quad\quad\quad\quad \int_N \mathbf{v}(\ln \rho)\mathcal{V} = -\int_N d*\mathbf{v} = -\int_V *\mathbf{v}.$

whenever *V* = ∂*N*.

If one interprets the functional *F*[*S*] as producing the *total magnetic charge* in an arbitrary compact orientable 2-dimensional submanifold *S* then one sees that if *S* bounds, so *S* = ∂*V*, then this makes physically intuitive sense and implies that the total magnetic charge in *V* is 0. However, when *S* does not bound a three-dimensional region, although the total magnetic charge in *S* might well be non-zero, nevertheless, one cannot physically interpret the integral as the magnetic charge contained in a three-dimensional region.

Of course, since we are defining a source current as a one-dimensional homology class and an electromagnetic field as a two-dimensional cohomology class one naturally must confront the question of whether there is only one such class or more than one, i.e., whether $H_1(M;\mathbb{R})$ or $H^2(M;\mathbb{R})$ vanishes. This naturally leads into the subject of our next section.

Finally, one must point out that there is still something formally incomplete about our system of axioms in that:

*a)* Ultimately it would be preferable to exhibit the fields *F* and *J* as having some intrinsic geometrical or topological origin, instead of introducing them as essentially logical primitives.

*b)* The constitutive law also seems to lack an immediate relationship to either the geometry or topology of the spacetime manifold.

Hence, the primary goals of the present axioms should probably be those of:
*a)* Exhibiting the spacetime metric and wave motion as special case corollaries to the constitutive laws.
*b)* Constructing the constitutive law from physical first principles and fundamental physical processes, as one might in the physics of condensed matter.
*c)* Illuminating a path into nonlinear electrodynamics that might define deeper foundations for quantum electrodynamics.
*d)* Suggesting a promising Ansatz for exhibiting the fields *F* and *J* as being naturally associated with geometrical or topological structures associated with the *SL*(4) reduction of the bundle of linear frames on spacetime that is necessary in order to facilitate Poincaré duality.

**5. Gauge symmetries and conservation laws.** The reason for the use of the plural in the title of this section is that actually, we have *two* ways in which ambiguity can arise in the construction of our basic fields *F* and *J*. The first is the usual one that relates to the possibility that the closed 2-form *F* is also exact; i.e., whether there exists a potential 1-form *A* such that:

(5.1) $\qquad\qquad F = dA.$

There are two comments that must be made about this: First, whether or not such an *A* exists depends upon the vanishing of $H^2(M;\mathbb{R})$. If that module does not vanish then one can find only local — or at best partial — potential 1-forms. Second, if *z* is a closed 2-form then *A*+*z* is also a potential 1-form for *F*. This defines an equivalence class of 1-forms by the equivalence relation:

(5.2) $\qquad\qquad A \sim A' \text{ iff } A - A' \in Z^1(M; \mathbb{R}).$

One refers to the equivalence relation as *gauge equivalence* and the transformation:
(5.3) $\qquad\qquad A \mapsto A + z$

as a *gauge transformation of the second kind.* In order to describe what the gauge transformations of the first kind were, we need to account for the appearance of the usual *U*(1) symmetry.

Customarily, one exhibits the closed form *z* as an exact form *d*λ, which means that zero-form λ represents a sort of infinitesimal generator for the gauge transformation. Naturally, the question of whether global *z* exist that are not exact depends upon the vanishing of $H^1(M; \mathbb{R})$, which is essentially the question of whether *M* is simply connected. However, since the customary physical arguments are usually local in scope, the Poincaré lemma allows us to find a suitable zero-form λ at least locally. This shows that the set of infinitesimal gauge transformations of the second kind can be effectively parameterized by the 0-forms. This means that we can regard the vector space of 0-forms as the (Abelian) Lie algebra of the group of gauge transformations. This group will either be the Lie group of positive smooth functions on *M* under multiplication, or the Lie group of smooth functions from *M* into $S^1$, which can also be considered as *SO*(2) or *U*(1).

If we make the latter choice, and give λ the form (which is no loss of generality):

(5.4) $$\lambda = d(\ln g) = g^{-1}dg,$$

where *g*: *M* → *U*(1), then, if we introduce a slight redundancy, the transformation (5.3) takes the form:

(5.5) $$A \mapsto g^{-1}Ag + g^{-1}dg$$

that one finds for the transformation of the local representative of a *U*(1) connection 1-form. One then refers to the group *U*(1) as a gauge group (of internal symmetries) for the field *F* and the choice of *g* (or λ) as a choice of local *U*(1) *gauge* for *F*, since *F* is locally equivalent to the pair (*A*, λ). This set of associations also carries with it the interpretation of *F* = *dA* as the curvature 2-form that is associated with the connection 1-form *A*.

In order to complete this geometric picture, we need to account for the appearance of a *U*(1) principal bundle *P* → *M* on which the connection 1-form *A* would be defined; one generally refers to such a principal fibration as a *gauge structure* for the field theory. The traditional reasoning is that quantum mechanics introduces a *U*(1) phase factor into the description of any particle by wave functions, whether they are photons, whose position distributions are obtained from complex wave functions, or electrons, whose position distributions are obtained from spinor wave functions. However, just as introducing a metric into the theory of electromagnetism seems premature when the metric should be a consequence of the propagation of electromagnetic waves, it also seems somewhat premature to introduce traditional quantum mechanical arguments into a theory that – one hopes – might shed new light on the deeper nature of quantum mechanics to begin with. Consequently, the author has been pursuing the possibility that the gauge structure for electromagnetism is not something one introduces axiomatically and independently from the geometry and topology of spacetime – as we already did with the fields *F* and *J* themselves – but something that follows naturally from the demands of physical interpretation. In particular, obtaining the gauge structure as actually an *SO*(2)-reduction of the bundle of linear frames on *M* seems promising (see [**17**]). However, that discussion would take us far from the immediate concerns, so we simply accept the traditional argument, for now, and proceed.

If one forms the electromagnetic field Lagrangian:

(5.6) $$\mathcal{L} = \tfrac{1}{2} dA(\mathfrak{h}) + A(J)$$

one sees that as long as one assumes that the constitutive equation gives $\mathfrak{h}$ as a function of *F*, and not *A*, then the kinetic energy term is gauge invariant. Under a gauge transformation, the interaction term *A*(*J*) picks up a contribution of:

(5.7) $$d\lambda(J) = J\lambda.$$

since this does not always vanish, we see that the interaction term breaks the gauge invariance of $\mathcal{L}$. By Nöther's theorem, the *U*(1) internal symmetry of $\mathcal{L}$ is associated with a conserved current:

(5.8) $$J = \frac{\delta \mathcal{L}}{\delta A},$$

which means that the $U(1)$ gauge symmetry is equivalent to charge conservation.

A less-discussed form of gauge equivalence is the one that pertains to $J$ when one addresses the question of whether a *superpotential* $\mathfrak{h}$ exists for $J$; i.e., an $\mathfrak{h} \in H_2(M;\mathbb{R})$:

(5.9) $$J = \delta \mathfrak{h}.$$

By the homological Poincaré lemma such an $\mathfrak{h}$ always exists locally. Whether this is also possible globally is a matter of whether $H_1(M; \mathbb{R})$ vanishes.

By the Hurewicz isomorphism theorem, this comes down to whether $M$ is simply connected or not. If $M$ is not simply connected, we could pass to the simply connected covering manifold $\tilde{M}$, where one would always have the existence of a globally defined $\mathfrak{h}$. However, since a global section of the covering map $s: M \to \tilde{M}$ would have to represent a global orientation, when $M$ is not orientable $s$ will necessarily be undefined at some points of $M$. The same will then be true of the image of $J$ in $\mathfrak{X}(\tilde{M})$, namely $s_*J$, as well as the pull back of $\mathfrak{h}$ to $M$ by $s$, namely $s^*\mathfrak{h}$.

Once again, we see that $\mathfrak{h}$ is not uniquely defined. Since $\delta^2 = 0$, if $Z \in Z_2(M;\mathbb{R})$ is a co-closed 2-form then the 2-form that one obtains by the transformation:

(5.10) $$\mathfrak{h} \mapsto \mathfrak{h} + Z$$

is also a superpotential for $J$. This defines a different sort of gauge equivalence for superpotentials:

(5.11) $$\mathfrak{h} \sim \mathfrak{h}' \quad \text{iff} \quad \mathfrak{h} - \mathfrak{h}' \in Z_2(M;\mathbb{R}).$$

Instead of looking for a 0-form to locally represent these transformations, we look for a 3-vector field $\Lambda$ such that $Z = \delta \Lambda$. By Poincaré duality, $\Lambda$ corresponds to a 1-form $*\Lambda$. Hence, the space of infinitesimal superpotential gauge transformations is parameterized by $\Lambda^1(M)$ this time. If we are only being local, we can exhibit these transformations by smooth functions from $M$ to $\mathbb{R}^4$. If we wish to regard this as the Abelian Lie algebra of some Lie group of functions from $M$ to an Abelian Lie group $G$ then one should observe that $\mathbb{R}^4$ can cover any of the Lie groups $\mathbb{R}^4$, $S^1 \times \mathbb{R}^3$, $T^2 \times \mathbb{R}^2$, $T^3 \times \mathbb{R}$, $T^4$. This level of ambiguity suggests that we are perhaps following the wrong ansatz.

Rather than pursue the analogous reasoning that we followed for the aforementioned gauge transformations, let us look at equation (4.13) again. This time, we shall put it into the form:

(5.12) $$\delta \mathbf{v} = i_\mathbf{v}(\rho^{-1} d\rho).$$

The 1-form $\rho^{-1} d\rho$ takes the same form as the $g^{-1} dg$ term that appeared in the context of the gauge transformations of $A$. We are then tempted to consider it to be a 1-form with values in the Lie algebra $\mathbb{R}$. However, since we are dealing with vector fields with non-zero divergence, we suspect that this time $\mathbb{R}$ is not the imaginary axis that generates the two-dimensional rotations, but the real axis that generates the dilatations, which is the multiplicative group $\mathbb{R}^+$.

If we recall that ρ originated when we chose to decompose $J$, which is assumed to have zero divergence, into ρ**v** then one could also say that, in effect, since this choice of decomposition for $J$ depends upon the choice of ρ, one must choose a *dilatational* gauge in order to define the vector field **v**. It is intriguing that such a dilatational gauge is also associated with a choice of parameterization for an integral curve of **v**, which also relates to mass and kinetic energy. At the moment, though, our concern is how it relates to charge conservation.

Now let us assume that $\rho^{-1}d\rho$ is the 1-form that appears when one makes a change of local gauge on an $\mathbb{R}^+$–principal bundle by way of ρ and transforms a local connection form σ from one gauge to the other:

$$(5.13) \qquad \omega \mapsto \rho^{-1}\sigma\rho + \rho^{-1}d\rho.$$

This time, we have no problem identifying the $\mathbb{R}^+$–principal bundle in question, since it is undoubtedly the associated $\mathbb{R}^+$–principal bundle to $L(S)$, which consists of all positive scalar multiples of the vector $J_x$ at any point $x \in M$. We shall denote this $\mathbb{R}^+$–principal bundle by $SL(S)$. Ordinarily, a line field on a manifold only defines an $\mathbb{R}^*$ (= $\mathbb{R} - \{0\}$)-principal bundle by way of the complement of its zero section. In order to reduce this to an $\mathbb{R}^+$–principal bundle, we would need to first show that the line field is orientable and then choose an orientation, which is simply a non-zero vector field on the support of the line field. However, this is also how we defined $L(S)$ in the first place, so the orientation is automatic. One must note that $SL(S)$ is topologically uninteresting, since the existence of a global section (over S) makes it trivial: $SL(S) = S \times \mathbb{R}^+$. Hence, an oriented one-frame on S is equivalent to smooth function on S with values in $\mathbb{R}^+$. In particular, if **v** = λ$J$ with λ>0 then the function associated with the oriented 1-frame **v** is simply λ.

The question then becomes that of how we physically account for the appearance of a connection 1-form on $SL(S)$. To address this issue, we first consider the role that it plays. Since ρ seems to be related to the scaling of the vector field $J$ within the oriented line field that it spans, the issue would seem to be how one compares the unit of (one-dimensional) "volume" in the fiber $L_x(S)$ at one point with the unit of volume at a neighboring point. (Of course, in dimension one a volume serves the same purpose as a norm or metric.) One must remember that in the absence of a metric there is nothing to distinguish any 1-frame of $L(S)$ as having a "unit" volume, except when one makes such an association arbitrarily. Hence, if that choice is to be truly arbitrary then the scaling factor that takes any 1-frame to a rescaled one must represent an internal symmetry of the field theory. However, one notes that in the conventional theory of electromagnetism this homothety invariance is only true for the electromagnetic waves themselves and is broken by the introduction of a mass term.

The connection 1-form ω that we defined takes its values in $\mathbb{R}$, which is the Lie algebra of $\mathbb{R}^+$. Like the electromagnetic potential 1-form, it too is an ordinary differential form. Consequently, we suspect that the only geometrical object that relates to it that also has an unambiguous physical meaning is its curvature, or field strength, 2-form:

(5.14) $$\Omega = d\omega.$$

(The term $\omega \wedge \omega$ vanishes because the Lie algebra of $\mathbb{R}^+$ is Abelian.)

Although it is tempting to speculate on whether $\Omega$ might relate, at least indirectly to the definition of the 2-form $F$, keep in mind that $\Omega$ is defined only over S, whereas $F$ is essentially defined on the complement of S. Hence, one would have to replace the definition of $F$ with a further "constitutive map":

(5.15) $$\Lambda^2(S) \to \Lambda^2(M-S).$$

In effect, this is what we usually get from solving the field equations: a map from the source to the field. Of course, since we are defining $\Omega$ to be an exact form and $F$ to be a closed form, if we pass to cohomology and the induced map is linear it would only take $[\Omega]$ to 0, so unless $[F] = 0$, as well – i.e., $F = dA$ – we should not expect this Ansatz to be productive in the context of topological electromagnetism unless the definition of $\Omega$ were weakened to something closed, but not exact.

**6. The constitutive axiom.** Naturally, the most open-ended axiom in the aforementioned set is the third one. The problem is to leave open the possibility that the constitutive law is nonlinear in such a way that it still relates to the geometrical or topological nature of the basic objects. One must unavoidably consider nonlinear theories of electromagnetism for both practical considerations, such as nonlinear optics, and more theoretical considerations, as when one is considering quantum electrodynamics. In either case, the domain of relevance tends to be the realm of large field strengths, such as the fields of high-energy laser beams, or the field of an electron or an atomic nucleus when one is sufficiently close.

The existence of vacuum polarization in QED tends to suggest a nonlinear constitutive law for the spacetime vacuum itself. (For a lengthier discussion of such matters, see [**18**].) Since the constitutive laws of macroscopic media are usually obtained by considering the macrostates (in the thermodynamic sense) that follow from the microstates of more fundamental processes, such as the formation of electric or magnetic dipoles in charge distributions, this seems to be the most promising direction for trying to probe the nature of the constitutive law that governs the electromagnetic vacuum state.

In a topological theory of electromagnetism, one might also wish that the microstates in question contain fundamental topological information. For instance, Misner and Wheeler [**19**] once attributed the appearance of charge itself to the existence of the much-discussed "spacetime wormholes," which amount to topological objects that render $H^2(M;\mathbb{R})$ non-trivial. An unacceptable aspect of the attachment of wormholes in the eyes of pre-metric electromagnetism is that they are three-dimensional objects that look like $S^1 \times S^2$ and are attached (by connected sum) to the three-dimensional spacelike leaves of a chosen proper time simultaneity foliation. In fact, for the purposes of most spacetime models, if one regards spacetime as the Cauchy development of an initial spatial manifold

Σ then the foliation is the "cylindrical" foliation $M = \mathbb{R} \times \Sigma$. In the absence of such a foliation, one can only attach four-dimensional objects to a four-dimensional manifold by connected sum and still produce a manifold. If one attaches, say, a point or a curve segment and wishes to still produce a four-dimensional manifold, one must make some further identifications and pass to a quotient, such as attaching the point at infinity to a plane by stereographic projection. Later, we shall discuss topological matters in the context of constitutive laws.

As we have defined the term, a constitutive law will generally be a differential equation in the two fields $J$ and $F$. However, as pointed out above, this equation may factor through a map $\mathfrak{h}: \Lambda^2(M) \to \Lambda_2(M)$ and the codifferential $\delta$ – i.e., $\delta\mathfrak{h} = J$ – which is the traditional form.

One can appeal to the theory of gravitation for an analogy. The spirit of Sakharov's notion of gravitation as "metrical elasticity" is that if spacetime curvature is essentially related to the second covariant derivative of the strain that is associated with deforming the Minkowski metric $\eta_{\mu\nu}$ into the Lorentzian metric $g_{\mu\nu}$ – à la the Cauchy-Green conception of strain – and this is coupled to stress-energy-momentum tensor then the Einstein field equations take the form of a constitutive equation in differential form.

One can give the constitutive axiom different forms, depending upon whether one assumes that $F = dA$ for some $A \in \Lambda^1(M; \mathbb{R})$ or $J = \delta\mathfrak{h}$ for some $\mathfrak{h} \in \Lambda_2(M; \mathbb{R})$. For instance, in the former case, one might consider laws of the form:

(6.1) $\qquad A = A(J), \qquad J = J(A), \qquad f(A, J) = \text{const.}$

In the latter case, one might consider laws of the form:

(6.2) $\qquad F = F(\mathfrak{h}), \qquad \mathfrak{h} = \mathfrak{h}(F), \qquad f(F, \mathfrak{h}) = \text{const.}$

In (6.1), the last expression represents a common form for the interaction term $A \wedge *J$ in the electromagnetic Lagrangian that expresses how the source current gets coupled to the electromagnetic field. Similarly, the last expression of (6.2) is of the form of the field kinetic energy term $F \wedge *H$ in such a Lagrangian. Of course, one usually needs a metric to form these expressions in order to define the 2-form $H$.

However, even without a metric, one can still combine the 2-form $F$ and the 2-vector field $\mathfrak{h}$, as well as the 1-form $A$ and the vector field $J$. This is due to the fact that 1-forms and 2-forms act as linear functionals on vector fields and 2-vector fields, respectively. Hence, one can form the expressions:

(6.3) $\qquad\qquad F \wedge *\mathfrak{h} = F(\mathfrak{h})\mathcal{V}, \qquad A \wedge *J = A(J)\mathcal{V},$

although these expressions are actually equivalent to forming:

(6.4) $\qquad\qquad F \wedge H, \qquad\qquad A \wedge J,$

when one deals with $H$ as a 2-form and $J$ as a 3-form. Later, we shall examine the first expression in the context of the intersection form of the spacetime manifold.

In local components, and neglecting the factor of $\mathcal{V}$, (6.3) gives the usual expressions:

(6.5) $\qquad\qquad \mathcal{L}_{\text{kin}} = \tfrac{1}{4} F_{\mu\nu} \mathfrak{h}^{\mu\nu}, \qquad \mathcal{L}_{\text{int}} = A_\mu J^\mu.$

On the Axioms of Topological Electromagnetism          21The 2-vector field $\mathfrak{h}$ of electromagnetic induction is the one that appears in the Maxwell equations, when we give them the form:

(6.6) $\qquad dF = 0, \qquad \delta\mathfrak{h} = J, \qquad \mathfrak{h} = \mathfrak{h}(F)$,

The homological Poincaré lemma says that one can, at least locally, find a 2-vector field $\mathfrak{h}$ such that:

(6.7) $\qquad\qquad\qquad J = \delta\mathfrak{h}$.

Of course, in the present case, as opposed to the usual machinery of Maxwell's equations in terms of differential forms, $\mathfrak{h}$ is a 2-vector field, instead of a 2-form. However, this was always implicit when one wrote the Maxwell equations in the older local – but still metric-free – form:

(6.8) $\qquad \partial_\mu F_{\nu\lambda} + \partial_\nu F_{\lambda\mu} + \partial_\lambda F_{\mu\nu} = 0, \qquad \partial_\mu \mathfrak{h}^{\mu\nu} = J^\nu, \qquad \mathfrak{h}^{\mu\nu} = \mathfrak{h}^{\mu\nu}(F_{\mu\nu})$.

The situation that is associated with the existence of a global $\mathfrak{h}$ on a covering manifold is distinct from the situation that relates to the existence of a global potential 1-form $A$ such that $F = dA$. In the former case, where the issue is the vanishing of $H_1(M; \mathbb{R})$, one can pass to the simply connected covering manifold of $M$. In the latter scenario, the issue is the vanishing of $H^2(M; \mathbb{R})$, which cannot be sidestepped by passing to a covering manifold, and leads to the possible existence of magnetic monopoles and spacetime wormholes.

If one is given the Lagrangian $\mathcal{L} = \mathcal{V}(F, J)$ a *priori* then one can retrieve the constitutive law from:

(6.9) $\qquad\qquad\qquad \mathfrak{h} = \dfrac{\partial \mathcal{L}}{\partial F}$.

If we sense that we are close to dealing with a spacetime metric then that is due to the fact that we are concerned with a kinetic energy, and, as one knows from relativistic point mechanics, kinetic energy is, in its simplest form, conformally related to the spacetime metric. Furthermore, the conformal factor is simply the square of the rest mass of the point particle; of course, in the case of electromagnetic waves this conformal factor would be zero.

Another piece of the geometric puzzle that is associated with constitutive laws comes from the fact that the speed of propagation of electromagnetic waves in vacuo – i.e., $c_0$ – is actually a derived quantity that follows from the assumed form of the constitutive law for the classical electromagnetic vacuum. That law amounts to the assumption that the electric and magnetic polarizability of that state is expressed by two constants, $\varepsilon_0$ and $\mu_0$, which give ([5]):

(6.10) $\qquad\qquad\qquad c_0^2 = \dfrac{1}{\varepsilon_0 \mu_0}$.

---

[5] Although the units in which $c_0 = 1$ are sometimes called "God's units," the fact that they tend to obscure the deeper truth of this physical subtlety suggests that they are probably "Satan's units," since God does not play games with truth.

As is well known, one of the most far-reaching consequences of quantum electrodynamics is the suggestion that this aforementioned assumption is an oversimplification of the nature of the electromagnetic vacuum state.  In particular, vacuum polarization seems to contribute at every turn to the nature of elementary electromagnetic processes.  Consequently, one expects that if classical electromagnetism leads to gravitation somehow then quantum electrodynamics might lead to quantum gravity (or at least an unambiguous definition of the concept).  Alternatively, if that is not the case then perhaps it is only quantum electrodynamics that leads to gravitation in the first place; indeed, this is probably more likely.

An intriguing hint of how this might come to pass is given by the fact that (6.10) also suggests that one can rewrite the simplest spacetime metric – i.e., the Minkowski space metric – in the form:

(6.11) $$\eta_{\mu\nu} = \text{diag}(\frac{1}{\varepsilon_0 \mu_0}, -1, -1, -1).$$

In this one elementary geometrical object, we can see Ansätze that might lead to the unification of gravitation, electromagnetism, vacuum polarization, and wave mechanics.

In order to gain more intuition about the nature of the problem of how the electromagnetic constitutive model leads to the spacetime metric as a consequence, we examine two special cases: the linear case and the nonlinear case as it relates to nonlinear optics and the Born-Infeld model, which is also an effective model for obtaining the vacuum polarization effects that were obtained by Heisenberg and Euler in the context of QED as consequences of a nonlinear electrodynamical theory. We then make some observations and speculations on the problem of resolving the constitutive laws into geometrical and topological aspects of more fundamental processes.

*a. Linear case.* In conventional linear electrodynamics one ordinarily assumes the *a priori* existence of a Lorentzian metric $g$, so the constitutive relation takes the 2-form $F$ to the 2-form $H$ by way of:

(6.12) $$H(F) = (\iota_g \times \iota_g) \bullet \chi(F),$$

where:

(6.13) $$\iota_g: T(M) \to T^*(M), \quad \mathbf{v} \mapsto i_\mathbf{v} g$$

is the linear isomorphism that takes each tangent vector to its metric-dual covector and:

(6.14) $$\chi: \Lambda^2(M) \to \Lambda_2(M), \quad F \mapsto \mathfrak{h} = \chi(F)$$

is a linear isomorphism of 2-forms with 2-vector fields.  In this form, we see that in an unpolarized medium the correspondence (6.12) is simply the identity map and one should have simply:

(6.15) $$\chi = (\iota_g \times \iota_g)^{-1}.$$

We then see that the effect of polarization is to deform the isomorphism of $\Lambda^2(M)$ with $\Lambda_2(M)$ that is defined when one has a Lorentzian metric $g$.  One might even consider the possibility that this deformation originates in a deformation of the metric itself, $g \mapsto g' = g + \delta g$, which should make:

$$\text{(6.16)} \qquad \iota_{g'} \times \iota_{g'} = \iota_g \times \iota_g + \iota_{\delta g} \times \iota_g + \iota_g \times \iota_{\delta g} + \iota_{\delta g} \times \iota_{\delta g}.$$

We could then define the polarization tensor field $\mathfrak{p}$ to be the difference between the deformed and undeformed states:

$$\text{(6.17)} \qquad \mathfrak{p} = (\iota_{g'} \times \iota_{g'})^{-1} - (\iota_g \times \iota_g)^{-1}.$$

To relate this to the usual definition of the polarization tensor field we define:

$$\text{(6.18)} \qquad P = (\iota_g \times \iota_g)\mathfrak{p} = (\iota_g \times \iota_g)(\iota_{g'} \times \iota_{g'})^{-1} - I,$$

which agrees with the usual conception of the polarization tensor field as $H - I$ if we set ([6]):

$$\text{(6.19)} \qquad H = (\iota_g \times \iota_g)(\iota_{g'} \times \iota_{g'})^{-1} = (\iota_g \times \iota_g)\chi.$$

Relative to a natural frame field, (6.12) takes the component form:

$$\text{(6.20)} \qquad H_{\mu\nu} = g_{\mu\rho} g_{\nu\sigma} \mathfrak{h}^{\rho\sigma} = g_{\mu\rho} g_{\nu\sigma} \chi^{\rho\sigma\tau\upsilon} F_{\tau\upsilon}.$$

We can then represent the matrix of the map $g \wedge g = (\iota_g \times \iota_g)$ by way of:

$$\text{(6.21)} \qquad (g \wedge g)_{\mu\rho\nu\sigma} = \tfrac{1}{2}(g_{\mu\rho} g_{\nu\sigma} - g_{\nu\rho} g_{\mu\sigma}).$$

When we are dealing with an oriented Lorentzian manifold, in addition to the isomorphism of $\Lambda^2(M)$ with itself that is defined by $(\iota_g \times \iota_g) \bullet \chi$, we also get another isomorphism from (the inverse of) Poincaré duality. If we compose the maps $\chi$ and $*$ then we get the isomorphism:

$$\text{(6.22)} \qquad *_\chi: \Lambda_2(M) \to \Lambda_2(M), \quad F \mapsto *\chi F.$$

We must be careful to distinguish the effect of this map from the previous one defined by $(\iota_g \times \iota_g) \bullet \chi$. In the limit of an unpolarized medium $\chi$ goes to $(\iota_g \times \iota_g)^{-1}$ so the map $(\iota_g \times \iota_g) \bullet \chi$ goes to $I$, but the map $*_\chi$ goes to the conventional form of the Hodge star isomorphism on a pseudo-Riemannian manifold. Hence, $*_\chi$ is deformation of the Hodge star isomorphism due to polarization.

We now wish to eliminate the explicit reference to the metric in the aforementioned constructions. We first represent the electromagnetic induction by the 2-vector field $\mathfrak{h} = \chi(F)$, instead of the 2-form $H$. Furthermore, we can alternatively represent the tensor field $\chi$ by the fully contravariant fourth rank tensor field of the form $\chi \in \Lambda_2(M) \otimes \Lambda_2(M)$ that corresponds to the linear isomorphism (6.14).

The components of a tensor field of the form $\chi \in \Lambda_2(M) \otimes \Lambda_2(M)$ come with two symmetries already: First, there is antisymmetry in the first and last pair due to the fact

---

[6] In this expression, we intend that $H$ be considered as an operator.

that they pertain to 2-forms. Second, from Lagrangian considerations, one often assumes that there is also a transposition symmetry of the form:

(6.22) $\qquad \chi(\alpha, \bullet) = \chi(\bullet, \alpha), \qquad \alpha \in \Lambda^2(M).$

This corresponds to a symmetry in the components of $\chi$:

(6.23) $\qquad \chi^{\mu\nu\tau\upsilon} = \chi^{\tau\upsilon\mu\nu}.$

This latter symmetry, combined with the fact that $\chi$ is a bilinear and non-degenerate functional on the vector bundle $\Lambda_2(M)$, implies that we can also regard $\chi$ as a fiber metric on the vector bundle $\Lambda_2(M)$. The question then arises: under what circumstances can we resolve the fiber metric $\chi$ on $\Lambda_2(M)$ to a Lorentzian one, $g$, on $\Lambda_1(M) = T(M)$? From (6.21), which we rewrite as $\chi = g \wedge g$, we see that we are essentially looking for the "exterior square root" of $\chi$.

From Poincaré duality, and the fact the 2-forms act as linear functionals on the 2-vector fields, one already has another scalar product defined for 2-forms on an orientable $M$:

(6.24) $\qquad <a, b> = {*}a(b) = i_{a \wedge b}\mathcal{V} = \mathcal{V}(a \wedge b).$

However, a simple comparison of symmetries shows that $\mathcal{V}$ cannot be represented by an exterior product of symmetric tensor fields, such as $g \wedge g$. Hence, if we wish to take the exterior square root of $\chi$ then we must subtract off the contribution proportional to $\mathcal{V}$:

(6.25) $\qquad \chi_0 = \chi - \alpha\mathcal{V},$

for some appropriate scalar $\alpha$.

In order to find an exterior square root of $\chi_0$, we follow Hehl, et al [**12, 13**] and define:

(6.26) $\qquad \# = {*}\chi_0 = {*}\chi + \alpha I,$

which is a linear isomorphism of $\Lambda^2(M)$ with itself.

The square of # is:

(6.27) $\qquad \#^2 = {*}\chi_0 {*}\chi_0.$

If we can find a form for $\chi_0$ that satisfies the constraint:

(6.28) $\qquad \#^2 = -1$

then we expect that we have reproduced the Hodge isomorphism, at least for 2-forms.

In order to solve (6.28) for $\chi_0$, at least locally, we choose a local frame field $\mathbf{e}_\mu$ whose domain is an open subset $U \subset M$ and whose reciprocal coframe field is $\theta^\mu$. We enumerate the basis for $\Lambda^2(U)$ defined by all $\theta^\mu \wedge \theta^\nu$ with $\mu < \nu$ by $E^I$, $I = 1, 2, \ldots, 6$. We can express the matrices of * and $\chi_0$ as block matrices:

(6.29) $$[*]_{IJ} = \begin{bmatrix} 0 & I \\ I & 0 \end{bmatrix}, \qquad [\chi_0]^{IJ} = \begin{bmatrix} A & C \\ C^T & B \end{bmatrix}$$

in which $A$ and $B$ are symmetric 3×3 real matrices. For the general solution of (6.28) $A$ and $C$ must take the form:

(6.30) $$A = pB^{-1} - \frac{1}{\det B} N, \qquad C = B^{-1} K$$

in which $K$ is an arbitrary antisymmetric 3×3 real matrix, which we express as $ad(k)$ for a $k \in \mathbb{R}^3$, where $ad$ refers to the Lie algebra on $\mathbb{R}^3$ that is defined by cross product, and:

(6.31) $$N = k \otimes k, \qquad p = \frac{tr(NB)}{\det B} - 1.$$

If one has a tensorfield $\chi_0$ that satisfies (6.28) then one can then decompose $\Lambda_2(M)$ into a direct sum $\Lambda_{+2}(M) \oplus \Lambda_{-2}(M)$ of three-dimensional eigenspaces of # with eigenvalues $\pm i$; one refers to these subspaces as the *self-dual* and *anti-self-dual* 2-forms, respectively, even though the eigenvalues are imaginary, not real, as in the Riemannian case.

Now, one can associate a conformal class of *Riemannian* metrics with every splitting of $\Lambda_2(M)$ into a direct sum $\Lambda_+(M) \oplus \Lambda_-(M)$ whose constituent sub-bundles have the same rank – viz., three. In fact, this corresponds to the splitting of Spin(4) into $SU(2) \times SU(2)$. It also relates to the fact the Hodge isomorphism is the same for all metrics in a conformal class.

Of course, the issue at hand is not to find a conformal class of Riemannian metrics, but a conformal class of *Lorentzian* metrics. As we have pointed out above, the existence of the former does not have to imply the existence of the latter. Similarly, the uniqueness of Riemannian metrics up to homotopy – they are all obtained by deformation retractions of the bundle of linear frames – does not imply the uniqueness of the homotopy class of Lorentzian metrics, which depends upon the vanishing of $H^3(M; \mathbb{Z}_2)$. Furthermore, the spin group that is associated with $SO(3, 1)$, namely, $SL(2; \mathbb{C})$, does not split the same way as Spin(4). However, if one regards a Lorentz boost $B$ as a "Wick-rotated" Euclidian rotation $R$:

(6.32) $$B = \mathfrak{w}^{-1} R \mathfrak{w}, \qquad \mathfrak{w} = \text{diag}(1, i, i, i)$$

then, at the infinitesimal level, the decomposition of the Lie algebra $\mathfrak{so}(4) = \mathfrak{so}(3) \oplus \mathfrak{so}(3)$ becomes the *vector space* decomposition of the Lie algebra $\mathfrak{so}(3,1) = \mathfrak{so}(3) \oplus \mathfrak{b}(3)$, where the vector subspace:

(6.33) $$\mathfrak{b}(3) = \mathfrak{w}^{-1} \mathfrak{so}(3) \mathfrak{w}$$

of infinitesimal boosts is not actually a Lie subalgebra of $\mathfrak{so}(3,1)$. The next question is whether we also have:

(6.34) $$\mathfrak{sl}(2;\mathbb{C}) \cong \mathfrak{su}(2) \oplus \mathfrak{b}'(2),$$

where $\mathfrak{b}'(2)$ is a three-dimensional subspace of infinitesimal boosts that one can obtain from $\mathfrak{su}(2)$ by a linear isomorphism. The answer is straightforward: By Hermitian polarization, we can decompose any $a \in \mathfrak{sl}(2;\mathbb{C})$ into a sum of a traceless Hermitian matrix and a traceless skew-Hermitian matrix:

(6.35) $$a = h^+ + h^- = \tfrac{1}{2}(h + h^\dagger) + \tfrac{1}{2}(h - h^\dagger).$$

Since the traceless skew-Hermitian part is an element of the Lie algebra $\mathfrak{su}(2)$, we need only verify that the traceless Hermitian part represents an infinitesimal boost and can be obtained from a traceless skew-Hermitian matrix by a linear isomorphism. The first part follows from the isomorphism of $\mathfrak{sl}(2;\mathbb{C})$ with $\mathfrak{so}(3,1) = \mathfrak{so}(3) \oplus \mathfrak{b}(3)$. As for the second part, the usual way to associate a Hermitian matrix with a skew-Hermitian one is to take $h^-$ to $-ih^-$. We cannot actually represent this by a conjugation, so we simply regard the transformation defined by multiplying Hermitian matrices by $i$ as the $\mathfrak{sl}(2;\mathbb{C})$ equivalent of the Wick rotation. We represent this situation by the commutative diagram:

$$\begin{array}{ccc}
& I \oplus \mathrm{Ad}(\mathfrak{w}) & \\
\mathfrak{so}(3) \oplus \mathfrak{so}(3) & \longrightarrow & \mathfrak{so}(3) \oplus \mathfrak{b}(3) \\
\downarrow & & \downarrow \\
& I \oplus -i & \\
\mathfrak{su}(2) \oplus \mathfrak{su}(2) & \longrightarrow & \mathfrak{su}(2) \oplus \mathfrak{b}(2)
\end{array}$$

in which the vertical arrows are the complexifications.

Hence, we conclude that a splitting $\Lambda_2(M) = \Lambda_{2+}(M) \oplus \Lambda_{2-}(M)$ is associated with a conformal class $[g]$ of Lorentzian pseudo-metrics, up to topological obstructions. (For the actual construction of a representative $g$ from $P$, see Hehl, et al.) This is equivalent to a global field of light cones on $M$.

In summation, we see that in the case of linear constitutive laws the assumption of symmetry in the tensor $\chi$, when regarded as a non-degenerate bilinear form on $\Lambda_2(M)$, defines $\chi$ as a fiber metric on that vector bundle. This, in turn, gives a splitting $\Lambda_2(M) = \Lambda_{2+}(M) \oplus \Lambda_{2-}(M)$ defined by the $\pm i$ eigenspaces of $\# = {}^*\chi_0$, and if the existence of a global Lorentzian metric on $M$ is topologically unobstructed then this leads to a conformal class $[g]$ of Lorentzian pseudo-metrics. Otherwise, when $M$ is compact one will have to restrict one's consideration to an open submanifold of $M$ that represents the complement of a finite set of points at which $[g]$ is undefined. (If $M$ is not compact then there is no topological obstruction to a global Lorentzian metric.)

Before we leave the realm of linear constitutive laws, we pause to point out another mathematically intriguing aspect of constitutive laws as scalar products on $\Lambda_2(M)$: Because the linear automorphism on $\Lambda_2(M)$ that is defined by $\# = {}^*\chi_0$ has the property that $\#^2 = -1$, it defines an *almost-complex* structure on $\Lambda_2(M)$. In effect, the splitting $\Lambda_2(M) = \Lambda_{2+}(M) \oplus \Lambda_{2-}(M)$ into self-dual and anti-self-dual 2-forms is analogous to their splitting into real and imaginary parts. Since the vector bundle $\Lambda_2(M)$ is of rank six as a real vector bundle, it will be of rank three as a complex vector bundle. This implies that its structure group is $GL(3; \mathbb{C})$. If one introduces an orientation and a Hermitian structure on $\Lambda_2(M)$ then one can reduce its structure group to $SU(3)$. In fact, the orientation on $T(M)$ implies an orientation on $\Lambda_2(M)$, and the scalar product $\chi$ defines a Hermitian structure by way of $\chi(\alpha, \beta^\dagger)$ where $\beta^\dagger = \text{Re}(\beta) - \text{Im}(\beta)$, in which $\beta = \text{Re}(\beta) + \text{Im}(\beta)$ is the decomposition of $\beta$ into self-dual and anti-self-dual parts. All of this might suggest a possible geometrical representation for the color $SU(3)$ gauge structure of the strong interaction without the necessity of representing $M$ as a complex three-manifold. However, our immediate concern is electromagnetism, so we let that pass.

*b. Nonlinear constitutive laws.* We might extend our constitutive law to something that is homotopic to linear; i.e., something of the form:

(6.36) $$\chi(F) = \chi^{(1)}(F) + N_s(F)$$

in which $\chi^{(1)}: \Lambda^2(M) \to \Lambda_2(M)$ is linear, $N_s: \Lambda^2(M) \to \Lambda_2(M)$ is nonlinear with $s \in [0, 1]$, and:

(6.37) $$\lim_{s \to 0} N_s = 0.$$

In such a case, we are dealing with a constitutive law of the "generalized Taylor series" form: linear map + (nonlinear map that vanishes in the weak field limit). In nonlinear optics [**20,21**], this level of approximation is referred to as "weak nonlinearity." The next levels of approximation are the quadratic and cubic ones ([7]):

(6.38a) $$\chi(F) = \chi^{(0)} + \chi^{(1)}(F) + \chi_s^{(2)}(F \odot F)$$
(6.38b) $$\chi(F) = \chi^{(0)} + \chi^{(1)}(F) + \chi_s^{(2)}(F \odot F) + \chi_s^{(3)}(F \odot F \odot F).$$

The leading constant term $\chi^{(0)}$ subsumes any residual polarization of the medium that remains when there is no applied electromagnetic field. Although that sounds physically uninteresting in the case of the classical spacetime electromagnetic vacuum, one should keep in mind that the Casimir effect is indicative of the reality of the zero-point field that one expects from quantum electrodynamic considerations. If one takes the position that quantum electrodynamics is really a phenomenological process for constructing the extension of linear electrodynamics into nonlinear electrodynamics in the realm of strong field strengths (see the "neoclassical" discussion in [**18**] or the quantum discussion in [**22**]) then one might still give this term serious consideration even for the electromagnetic vacuum.

The terms $\chi_s^{(2)}$ and $\chi_s^{(3)}$ are linear maps from the respective symmetrized tensor products of $\Lambda^2(M)$ with itself to $\Lambda_2(M)$. Hence, they can be represented in either form:

(6.39a) $$\chi_s^{(2)}: \Lambda^2(M) \odot \Lambda^2(M) \to \Lambda_2(M),$$
or: $$\chi_s^{(2)} \in \Lambda_2(M) \odot \Lambda_2(M) \otimes \Lambda_2(M),$$

---

[7] The notation $\odot$ refers to the symmetrized tensor product.

(6.39b) $$\chi_s^{(3)}: \Lambda^2(M) \odot \Lambda^2(M) \odot \Lambda^2(M) \to \Lambda_2(M),$$

or: $$\chi_s^{(3)} \in \Lambda_2(M) \odot \Lambda_2(M) \odot \Lambda^2(M) \otimes \Lambda_2(M).$$

Furthermore, $\chi_s^{(2)}$ and $\chi_s^{(3)}$ are expected to vanish in the limit $s \to 0$, which one intends to be the limit of weak field strength. Hence, $s$ represents a sort of coupling constant for the nonlinear terms.

Although it would be mathematically natural to generalize all of this to the methodology of jet bundles or covariant derivatives in order to account for the successive terms as true higher-order derivatives of $\mathfrak{h}$ with respect to $F$, we shall take a more nonlinear optical route and treat the successive terms as simply phenomenologically defined tensor fields. Of course, the ultimate challenge to theoretical physics is to account for these tensor fields in terms of geometrically or topologically defined tensor fields; i.e., in terms of spacetime structure. Consequently, one should not keep the aforementioned generalities out of one's consideration completely.

In nonlinear optics, since([8]):

(6.40) $$H(F) = F + P(F) = (I + P)(F),$$

one correspondingly regards the polarization 2-form $P(F)$ as the essential part in the eyes of nonlinear analysis and then decomposes the nonlinear operator $P$ into linear, quadratic, etc., parts:

(6.41a) $$P(F) = P^{(0)} + \chi^{(1)}(F) + \chi^{(2)}(F \odot F)$$
(6.41b) $$P(F) = P^{(0)} + \chi^{(1)}(F) + \chi^{(2)}(F \odot F) + \chi^{(3)}(F \odot F \odot F).$$

However, since we are trying to avoid using the spacetime metric in the fundamental statements about electromagnetism, and $P$ seems to relate mostly to deformations of the metric, we shall simply regard (6.38a,b) as the approach we shall take to weak nonlinearity.

Of all of the effective theories of nonlinear electrodynamics that are based in quantum electrodynamical considerations the one that seems to get the most attention is the Born-Infeld theory [**15**]. It is based in the imposition of a maximal field strength that would precede the onset of vacuum polarization and actually agrees with the computations of Euler and Heisenberg [**23,24**] of the vacuum polarization tensor that would follow when one starts in quantum electrodynamics.

The Born-Infeld Lagrangian for the electromagnetic field $F$ takes the form:

(6.42) $$\mathscr{L} = (\sqrt{E_c^2 + \mathscr{F} - \mathscr{G}} - E_c)\mathscr{V},$$

in which $E_c$ represents a critical electric field strength.

In order to define the Born-Infeld Lagrangian, one must first introduce the field invariants:

(6.43) $$I_1 = F \wedge *F = \mathscr{F}\mathscr{V}, \qquad I_2 = F \wedge F = \mathscr{V}(F, F)\mathscr{V} = \mathscr{G}\mathscr{V}.$$

---
[8] To be truly faithful to nonlinear optics, we should be talking about the electric field by itself, not the Minkowski 2-form, since the effects of large magnetic field strengths are usually treated as a separate class of phenomena. However, since our optical medium is the spacetime vacuum, and the $E$-$B$ decomposition follows from the imposition of a metric and a time orientation, we shall be more cautious from the outset.

These correspond to the local expressions:

(6.44) $\quad\quad\quad \mathcal{F} = \tfrac{1}{2} g^{\mu\nu} g^{\rho\sigma} F_{\mu\nu} F_{\rho\sigma}, \quad\quad\quad \mathcal{G} = \varepsilon^{\mu\nu\rho\sigma} F_{\mu\nu} F_{\rho\sigma}.$

Ordinarily, one intends that the * isomorphism is due to Hodge duality, which, in turn, comes from the imposition of a Lorentzian structure on spacetime, which is why one usually constructs one's field Lagrangian from *Lorentz* invariant expressions in $F$ – à la Mie [**25**]. In the pre-metric case, one must look for *SL*(4) invariant expressions. Of the latter two expressions, only $I_2$ can be formed without introducing a metric, at least as stated. Indeed, it can be defined (as a 4-form) without recourse to $\mathcal{V}$, so it is actually *GL*(4)-invariant. The expression $\mathcal{F}$ takes the form of $\|F\|^2$ when one gives $\Lambda^2(M)$ the norm that comes from the scalar product $g^\wedge g$.

However, recall that if we are considering a $\chi$ such that $\#^2 = -1$, where $\# = {}^*\chi_0$ then, in effect, we have reconstructed the Hodge * star, which is equivalent to a conformal class of metrics. Consequently, given such a $\chi$ we can still use the # isomorphism it defines in place of the Hodge star, and define a generalization of $I_1$ by:

(6.45) $\quad\quad\quad I_1 = F \wedge \#F = \chi(F, F)\mathcal{V}.$

However, since we have defined our replacement for the Hodge star only on 2-forms, none of the remaining three Lorentz invariants that Mie (with a correction from Weyl and Born) obtained, namely:

(6.46a) $\quad\quad\quad I_3 = A \wedge {}^*A$
(6.46b) $\quad\quad\quad I_4 = i_A F \wedge {}^* i_A F$
(6.46c) $\quad\quad\quad I_5 = i_A {}^*F \wedge {}^* i_A {}^*F$

can be defined without using a metric, since they all take the form of norm-squares of 1-forms.

The Lagrangians that involve norm-squares using the spacetime metric usually represent kinetic energy terms for the field, just as the elementary expression for the non-relativistic kinetic energy of a point mass involves using a conformal transform of the spatial metric tensor, whose conformal factor is $1/2m$. Similarly, the potential energy in a one-dimensional (or isotropic three-dimensional) Hookean material that has been displaced $\Delta x$ from its equilibrium state involves a conformal transform of the spatial metric whose conformal factor is $1/2k$.

However, when one goes from point matter to extended matter, or isotropic to anisotropic media, or linear to nonlinear media, all of the aforementioned simplifications break down and the conformal transformation of the metric must be – in the linear case – replaced with a product of a conformal transformation and a volume-preserving shear, and with more elaborate expressions when one goes to the nonlinear realm.

In any event, the generalizations all amount to constitutive laws in one form or another. A metric tensor then emerges as essentially a special case of a process by which one associates tensor fields with their dual tensor fields. As we have seen, in the case of 2-forms on a four-manifold, a volume element gives another.

**7. Topological constitutive laws.** In order to make electromagnetism a completely topological phenomenon, in addition to giving a topological interpretation to the fundamental fields, we also need to find some topological origin for the constitutive properties of the medium in question. This is especially perplexing in the case of the spacetime vacuum itself, since, classically, the spacetime vacuum was assumed to be have an electromagnetic structure that was expressed only by the constants $\varepsilon_0$ and $\mu_0$. However, as pointed out above, quantum electrodynamics suggests that this picture of the spacetime vacuum is an oversimplification.

*a. The intersection form.* Since we are replacing our electromagnetic field strength 2-form $F$ with its de Rham cohomology class $[F]$, we might also consider a constitutive axiom in the form of a bilinear functional on the second de Rham cohomology space itself, instead of $\Lambda_2(M)$:

(7.1) $$H^2(M; \mathbb{R}) \times H^2(M; \mathbb{R}) \to \mathbb{R}.$$

Note that this time we are defining our bilinear functional on a *vector space*, not a $C^\infty(M)$-*module*, as we did for linear constitutive laws.

Now, any closed oriented orientable 4-manifold $M$ already has one such functional defined by Poincaré duality:

(7.2) $$Q([\alpha], [\beta]) = \int_M \alpha \wedge \beta = Q([\beta], [\alpha]).$$

Inside the integral, $\alpha$ and $\beta$ are any closed 2-forms that define each chosen de Rham class in dimension two. When $M$ has no boundary, the integral will be independent of this choice of representative. If $M$ has a boundary then, by Stokes's theorem, replacing $\alpha$ by $\alpha + d\lambda$ and $\beta$ by $\beta + d\mu$ will produce a boundary term of $\alpha \wedge \mu - \lambda \wedge \beta + d\lambda \wedge \mu$, which does not have to vanish, except for a restricted class of gauge transformations, such as $\lambda, \mu$ that vanish on $\partial M$.

What is not entirely obvious in (7.2) is how Poincaré duality and the choice of $\mathcal{V}$ affects the definition of $Q$. Hence, we rewrite it as:

(7.3) $$Q([\alpha], [\beta]) = \int_M \alpha (*\beta) \mathcal{V},$$

in which the role of our choice of $\mathcal{V}$ and its associated Poincaré duality isomorphism $*$ becomes clear.

$Q$ is not only non-degenerate, but unimodular, as well, so its symmetry and bilinearity imply that $Q$ defines a scalar product on the vector space $H^2(M; \mathbb{R})$. The bilinear functional $Q$ is called the *intersection form* for a closed orientable $M$ [**10**, **26-28**]. As we shall see shortly, it plays a fundamental role in the topology of four-dimensional differentiable manifolds.

The relevance of the intersection form to electromagnetism should be unavoidable in our present formulation, since we are defining electromagnetic fields to be two-dimensional de Rham cohomology classes to begin with. Of course, the only way that $Q$ is non-trivial is if we assume that the topology of $M$ has a non-vanishing $H^2(M; \mathbb{R})$, i.e., that some $F$'s do not admit global potential 1-forms, which gets one back into the realm of magnetic monopoles, wormholes, and other spacetime pathologies.  Hence, we briefly summarize some of the fundamental features of the role that the intersection form plays in the topology of four-manifolds.

  *b. The topology of four-manifolds*[**26-28**].  Algebraically, a scalar product can be classified by two integers: its *rank*, which equals the second Betti number for *M* in the present case:

(7.4) $$b_2 = \dim(H^2(M; \mathbb{R})),$$

and its *signature*:

(7.5) $$\tau = b^+ - b^-.$$

where $b^+$ is the maximal dimension of the subspaces of $H^2(M; \mathbb{R})$ on which $Q$ is positive definite and $b^-$ is the maximum dimension of the subspaces on which it is negative definite.

When one restricts the form $Q$ to the free part of the integer cohomology, i.e., the free part of $H^2(M; \mathbb{Z})$, which defines an integer lattice of the form $\mathbb{Z}^{b_2}$, one can also define the *type* of $Q$ according to whether it takes it values in the even integers or not.  In the affirmative case one calls $Q$ *even*, and in the contrary case one calls $Q$ *odd* (although that does not have to imply that *all* of its values are odd integers).

For indefinite forms, the Hasse-Minkowski classification says:

(7.6) $$Q = \begin{cases} I_l \oplus (-I_m) & \text{when } Q \text{ is odd} \\ l\begin{bmatrix} 0 & 1 \\ 1 & 0 \end{bmatrix} \oplus mE_8 & \text{when } Q \text{ is even.} \end{cases}$$

In this classification, the matrix $E_8$ is the Cartan matrix of the exceptional simple Lie algebra $E_8$.

For definite forms, the classification is not as straightforward.  However, as we shall discuss later, Donaldson's theorem narrows down the field of definite forms that describe the intersection forms of simply connected closed four-manifolds considerably.

For simply connected four-dimensional manifolds, the intersection form is fundamental to the other characteristic classes, since they can all be derived from the information that is contained in $Q$:

  Second Stiefel-Whitney: $Q(w_2[M], \alpha) = Q(\alpha, \alpha) \pmod 2$    for all $\alpha \in H^2(M; \mathbb{Z}_2)$
  First Pontrjagin ([9]):   $p_1[M] = 3\tau\, [\mathcal{V}] = 3(b^+ - b^-)[\mathcal{V}]$
  Euler:                    $e[M] = (b^+ + b^-)[\mathcal{V}]$.

(In these expressions, $\mathcal{V}$ is a choice of volume element.)  The first and third Stiefel-Whitney classes vanish by simple connectedness and Poincaré duality; indeed this is actually true of $w_3$ for any compact orientable 4-manifold.  The fourth Stiefel-Whitney class is the $\mathbb{Z}_2$ reduction of $e[M]$.

---

[9] This result follows from either the Hirzebruch signature theorem [**10, 29**], or, more generally, the Atiyah-singer index theorem [**29,30**].

The second Stiefel-Whitney class plays an important role in the context of Spin(4) structures on *M*, namely it must vanish in order for an *SO*(4)-reduction of *GL*(*M*) to admit a two-to-one covering by a Spin(4)-principal bundle. Moreover, the homotopy classes of such bundles are indexed by $H^1(M; \mathbb{Z}_2)$. In particular, if a simply connected four-manifold admits a spin structure then it must be unique. Moreover, a simply connected four-manifold admits a spin structure iff its intersection form, when restricted to integer cohomology classes, takes on only even values.

The mathematician's understanding of how the intersection form *Q* relates to the topology of simply connected four-manifolds is growing quite extensive by now. In 1949, Whitehead showed that if two closed simply connected four-manifolds have isomorphic intersection forms then they are homotopy equivalent. In fact, every unimodular symmetric bilinear form is the intersection form of some simply connected *topological* four-manifold. However, not all of them will be smoothable, or even piecewise-linear.

For the case of indefinite intersection forms, one has mostly examples to deal with. For instance, one can construct a compact orientable four-manifold with an intersection form diag(1, …, 1, −1, …, −1) = $l(1) \oplus m(-1)$ by taking the connected sum of *l* copies of $\mathbb{C}P^2$ (regarded as a real four-dimensional manifold instead of a complex two-dimensional one) with one orientation with *m* copies of $\overline{\mathbb{C}P^2}$, which is $\mathbb{C}P^2$ with the opposite orientation. This is because $H^2(\mathbb{C}P^2; \mathbb{Z}) = \mathbb{Z}$, reversing the orientation inverts the sign of the intersection form, and the connected sum operation produces a direct sum in the cohomology. We point out that the standard generator for $H^2(\mathbb{C}P^2; \mathbb{Z})$ is essentially $\mathbb{C}P^1$, which is diffeomorphic to a 2-sphere.

For the case of definite intersection forms, one has Donaldson's celebrated theorem that the only positive definite intersection form on a smooth closed simply connected four-manifold is the standard one.

Now that we have discussed the notion of the intersection form of an orientable manifold, we can examine one possibility for rendering $H2(M; R)$ non-trivial in a manner that seems closely related to the physical process of vacuum polarization in the presence of strong electric field strengths.

*c. A possible topological mechanism for vacuum polarization.* As a first attempt at giving a deeper physical basis for a topological constitutive axiom let us take the process of vacuum polarization literally and see what sort of expressions we can derive from it.

The basic process in vacuum polarization is that of the conversion of a photon into a virtual electron/positron pair when the field strengths of the photon exceed some threshold value. We first assume that the reason for using the word "virtual" is simply the fact that these processes and their inverses may be happening over such a short time interval that the measurement process cannot resolve them in experimental practice, even though their *collective* effect as a macroscopic ensemble is measurable; for instance, the Lamb shift. Hence, a virtual process might represent a real, but unobservable, process.

We assume that any sufficiently small region of the spacetime electromagnetic vacuum can exist in one of three phases:
*a)* Unpolarized; this is the *weak* field case.
*b)* Polarized by a bound electron/positron pair; this is the *critical* field case.

*c)* Stably charged, when the binding energy has been exceeded and the charges form stable disjoint distributions; we shall call this the *super-critical* field case.

Our main concern in this study is the transition between the first two phases, although we shall treat both phase transitions as if they were associated with the spontaneous breaking of the vacuum symmetry and the appearance of topological defects.

We next observe that when a photon resolves to a bound electron/positron pair, as long as the two charges are assumed to be separated by an actual displacement vector **d** there is also an electric dipole 2*e***d** that is associated with the pair. The vector **d** brings about a reduction in the symmetry of the vacuum state from the *SO*(2) gauge symmetry of the electromagnetic Lagrangian that represents charge conservation to the {*e*} symmetry of the vacuum state that a choice of non-zero vector brings about. The homogeneous space that is defined by this reduction is also *SO*(2) – or $S^1$, if you prefer – whose first non-trivial homotopy group is $\pi_1(S^1) = \mathbb{Z}$. Hence, one should expect the topological obstruction to making such a reduction to be an element of $H^2(M; \mathbb{Z})$. In the language of defects, we are associating a "line defect" to some 2-cycles in *M*, which we regard as belonging to the singularity complex of *F* – i.e., its sources. In the usual Dirac construction of magnetic monopoles, if the 2-cycle is a 2-sphere then this line defect is its equator.

Consequently, let us imagine that topologically the formation of a charge pair is equivalent to the cavitation of a bubble in a fluid, i.e., a 2-cycle that does not bound. Furthermore, we regard this topology-changing process as described by the contraction of an open 2-ball to a point, while leaving a 2-sphere as a boundary to the remaining space. This is reminiscent of the fact that bubbles cannot form in real fluids unless there is point on which to nucleate. Note that the point that remains when one contracts a 2-ball is no longer in the same connected component of the spacetime manifold ([10]). Hence, we have added one generator to $H^2(M; \mathbb{Z})$ and one generator to $H^0(M; \mathbb{Z})$. We could also regard this process as one of attaching a copy of $\mathbb{C}P^1$ by connected sum and a point by disjoint union, although we shall disregard the contribution that the point makes and concentrate on the remaining component of the spacetime manifold. Hence, this sort of topological modification produces an increase in the second Betti number of *M*, as well as the signature of *M*, so it should be manifest in the structure of *Q*.

We will assume, by symmetry, that the charge is effectively concentrated at the poles with +*e*, by definition, at the North pole and –*e* at the South pole. Hence, the polar diameter, which we orient from the positive charge to the negative one, defines the displacement **d**, as well as the line defect that gets associated with the 2-cycle that the bubble defines. As the electromagnetic field strength grows beyond the critical value for cavitation, one expects the 2-sphere to elongate into an ellipsoid of increasing polar diameter and decreasing equatorial diameter.

If one imagines this process of cavitation under high electric field strength as also happening in great numbers – which represents a somewhat different conception of "spacetime foam" – then one can pass to the continuum limit and define an overall density of electric dipoles; i.e., a polarization tensor field $\mathfrak{p}$, or simply the tensor field $\chi$.

---

[10] Indeed, it has the wrong dimension to be included in the manifold that it left behind.

Note that the second phase transition that one can envision under vacuum polarization, viz., the actual separation of the charge pair, when it goes from a bound state to a scattering state, also represents a further topology-changing process. The single bubble that cavitated under the action of the surrounding field has become two disjoint bubbles. As for the point that the open 3-ball contracted to, although it sounds trivial, nevertheless, in the eyes of topology something must be done about it. In particular, if there is true mirror equivalence between matter and antimatter, one expects that if one of the free states has a disjoint point "inside" it then so should the other. However, the presence of such a point might also be an indicator of the instability of the topological state. When one removes the external field, one expects the vacuum polarization to disappear, and one imagines the spherical polarized boundary that previously formed collapsing back into the central point. Since electrons and positrons presumably exist in the absence of external fields as well, one might expect that the central point no longer exists in either.

A possible accounting of where the point went to is provided by the fact that since the process of separating one bubble into two bubbles – like the process of forming a bubble in the first place – cannot be effected by any continuous map there must be singularities for that process, i.e., points at which the map is discontinuous. If one imagines the bubble being stretched along the polar axis under increasing field strengths until the equator starts "necking down" then it is conceivable that the central point now serves as a point to which the equator contracts in order for the hemispheres to separate into disjoint bubbles. We schematically illustrate the process in Fig. 1, in which the parameter that changes between phases is the increasing electric field strength.

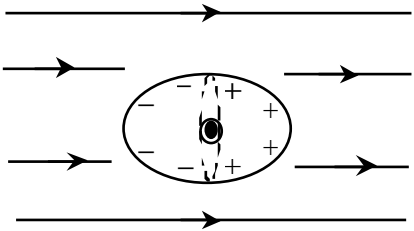

Fig. 1a. Formation of bubble

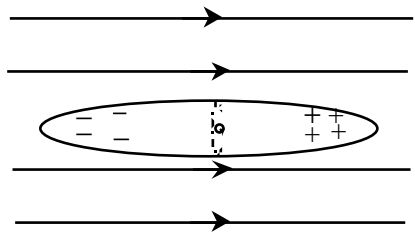

Fig. 1b. Elongation.

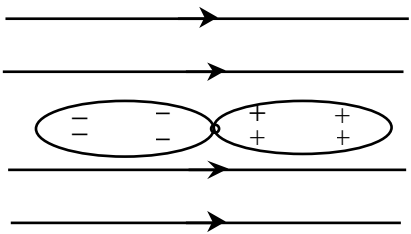

Fig. 1c. Constriction of equator.

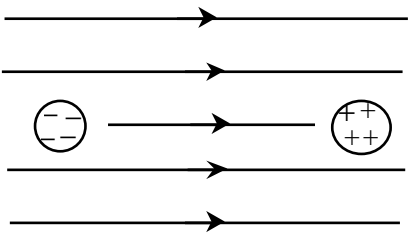

Fig. 1d. Separation of new bubbles.

One could regard this process as a higher-dimensional analogue of the two-dimensional Lorentz cobordism that takes the form of the "trouser manifold." In that process, one facilitates the subdivision of one circle (the "waistline") into two disjoint circles (the two "cuffs") by means of a two-dimensional Lorentzian manifold whose boundary consists of the oriented disjoint union of the three circles and whose line field describes the dynamical system that effects this transition. Of course, a compact two-dimensional manifold $M$ will admit a Lorentzian structure iff its Euler-Poincaré characteristic:

(7.7) $$\chi[M] = 2 - 2g$$

vanishes, in which we have introduced the integer $g$ that one calls the *genus* of $M$. Since this only vanishes for $g = 1$, we see that the trouser manifold ($g = 3$) does not admit a Lorentzian structure without singular points. But then, if one thinks of a Lorentzian structure as a one-dimensional differential system that describes the process of proper time evolution then one sees that the aforementioned process cannot be described by a flow of global diffeomorphisms anyway, or even a global flow of homeomorphisms.

In the case of bubble cavitation, we are looking at, first, the formation of a 2-sphere and a point, and then the separation of one 2-sphere into two disjoint 2-spheres in which the equator contracts to the point that was formed in the first transition and then gets absorbed into the surrounding manifold. If these three 2-spheres collectively define the boundary of a compact oriented 3-manifold then, since Poincaré duality makes $\chi[M] = 0$ for any compact orientable 3-manifold, that oriented cobordism can also be given a Lorentzian structure.

Since the super-critical phase transition has attached another copy of $\mathbb{C}P^2$ and thereby altered $Q$, a crucial issue is whether the change to the signature is reflected in the dimension of the maximal negative subspaces or that of the maximal positive subspaces. One suspects that an anti-particle should somehow balance a particle in a symmetrical sort of way, so perhaps the anti-particle has the opposite orientation and we are really attaching $\mathbb{C}P^2$ for the particle and $\overline{\mathbb{C}P^2}$ for the anti-particle. This would suggest that the symmetry that relates most fundamentally to the matter/anti-matter symmetry is the $\mathbb{Z}_2$ symmetry that one associates with the choice of orientation.

The possibility that past a critical electric field strength virtual or actual electron-positron pairs would form in such a way that the ultimate field strength would be reduced is usually referred to as the *Klein paradox*. If one regards such a process as simply another example of a more general class of phase transitions that are associated with spontaneous breaking of the ground state symmetry and the formation of topological defects then perhaps the phenomenon will not seem so paradoxical.

**8. Electromagnetic waves.** Since the appearance of the Minkowski scalar product in electromagnetism seems to be concerned with the appearance of wavelike solutions to the electromagnetic field equations, we should examine the way that one might use the structure of wave motion to induce the spacetime metric.

First, one might point out that the second order hyperbolic PDE that one usually considers to be the linear wave equation is not necessarily the most fundamental statement about the intrinsic nature of wave motion. For instance, one can also consider *first* order hyperbolic PDE's that one refers to as *conservation laws*. For instance, the two-dimensional linear wave operator can be regarded as the composition of two first order ones that define the equations:

(8.1) $$\partial_t \phi - c^{-1}\partial_x \phi = 0, \qquad \partial_t \phi + c^{-1}\partial_x \phi = 0,$$

that have solutions of the form:

(8.2) $$f(t, x) = f(t - c^{-1}x), \qquad g(t, x) = f(t + c^{-1}x).$$

Hence, just as the second order d'Alembertian operator is the product of two first order differential operators, similarly, the usual d'Alembert solution of the second order linear wave equation is simply the sum of solutions to two first order equations.

More generally, one can consider conservation laws of the form:

(8.3) $$\partial_t \phi + c(t, x, \phi)\partial_x \phi = 0,$$

which include many of the popular nonlinear wave equations, as well. When one generalizes (8.3) to $n+1$ dimensions so that $\partial_x \phi$ becomes an $n$-dimensional vector hyperbolicity becomes a property of the map $\chi$. Of course, a drawback to this formulation of wave equations is that is presumes that space is distinct from time.

In the context of exterior differential forms, the equivalent statement to (8.1) is that when a *k*-form $\alpha$ on an orientable Riemannian (Lorentzian, resp.) manifold is both closed and co-closed:

(8.4) $$d\alpha = 0, \qquad \delta\alpha = 0,$$

it is also harmonic:

(8.5) $$\Delta\alpha = 0, \qquad (\square\alpha = 0, \text{ resp.})$$

where:

(8.6) $$\Delta = \square = \delta d + d\delta.$$

In the Riemannian case, the converse is also true. The fact that the converse does not have to be true in the Lorentzian case, along with the fact that many of the methods of Hodge theory in general break down for Lorentzian manifolds, is exactly why we choose to distinguish the Laplace operator from the d'Alembertian operator.

In particular, the usual source-free Maxwell equations on a Lorentzian manifold:

(8.7) $$dF = 0, \qquad \delta F = 0$$

imply that $F$ is harmonic in the d'Alembertian sense – i.e., wavelike. In fact, the inclusion of a source current $J$ simply changes the homogeneous wave equation to a forced wave equation.

To bring the discussion back to the pre-metric context, we point out that if our manifold is at least orientable then we can make sense of the codifferential operator $\delta$ only when applied to $k$-vector fields. The compositions $\delta d$ and $d\delta$ then become absurd. Consequently, we might look for a way around the use of the metric.

For instance, one might replace the metric with the constitutive law defined by $\chi$, which still gives us a generalized Hodge duality − at least for 2-forms − by way of $\# = *\chi$. Hence, we define the predictable codifferential operator on 2-forms:

(8.8) $\qquad \delta_\chi: \Lambda^2(M) \to \Lambda^2(M), \qquad \delta_\chi \alpha = - \#d\#\alpha.$

However, in order to complete the definition of a wave operator associated with $\chi$ we also need to define such a codifferential operator on 3-forms. This would necessitate the extension of #, hence $\chi$, to 3-forms. By Poincaré duality this would also imply an isomorphism of $\Lambda^1(M)$ with $\mathfrak{X}(M)$, or $T^*(M)$ with $T(M)$. Since such an isomorphism might then, by symmetrization, define a metric on $M$ we suspect that perhaps the introduction of a metric is inevitable in order to define wave motion.

There are reasons to suspect that this situation is based in deeper considerations about the nature of wave motion. As has been discussed elsewhere [**17**], an essential construction that is associated with wave motion in a four-dimensional manifold is that of a pair of transverse foliations of $M$ of codimension one. One of them defines the proper simultaneity leaves relative to a choice of rest frame. The other defines the isophase hypersurfaces for the wave motion. Their intersections are two-dimensional instantaneous isophases.

Although we have been avoiding the introduction of the proper time simultaneity foliation as being essentially equivalent to the introduction of a metric, as it happens, we can still define the aforementioned two-dimensional foliation of instantaneous isophases purely on the basis of the algebraic properties of $F$.

The *rank* of a $k$-form $\alpha$ on a manifold $M$ is defined to be the maximum dimension of its annihilating subspace $A_x$ at each point $x \in M$. More precisely, a vector $\mathbf{v} \in T_x(M)$ is in $A_x$ iff $i_\mathbf{v} \alpha = 0$. Another way of characterizing the rank of an exterior $k$-form $\alpha$ is that it is the smallest integer $r$ such that $\alpha^r = 0$, in which we intend that the power refers to the exterior product.

For a 2-form, the rank must be even, so for a four-dimensional $M$, the rank of a 2-form can only be 0, 2, or 4. The latter possibility corresponds the case in which the 2-form is simply 0 itself. Hence, if the 2-form in question is $F$ then the physically interesting cases are when the 2-form in question has rank 2 or 4.

When $F$ has rank 4 four, there are linearly independent 1-forms $\alpha, \beta, \gamma, \delta$ such that:

(8.1) $\qquad F = \alpha \wedge \beta + \gamma \wedge \delta.$

When $F$ has rank 2, there are linearly independent 1-forms $\alpha$ and $\beta$ such that:

(8.2) $\qquad F = \alpha \wedge \beta.$

One refers to the set $\{\alpha, \beta\}$ or $\{\alpha, \beta, \gamma, \delta\}$ as an *associated system* to $F$.

Alternately, if $F$ has rank 2, one must have that:

(8.3) $\quad F \wedge F = 0.$

When one gives $F$ the usual $E$-$B$ decomposition, namely, $F = \theta \wedge E + *(\theta \wedge B)$ for some choice of timeline unit 1-form $\theta$, this last condition takes the form:

(8.4) $\quad E \wedge *B = g(E, B) = 0.$

Hence, if $F$ has rank two then the $E$ and $B$ fields, which are spacelike in conventional electromagnetism, must be orthogonal or zero.

The 1-forms $\alpha$ and $\beta$ collectively span a two-dimensional sub-bundle of $T^*(M)$ and annihilate a two-dimensional sub-bundle D of $T(M)$. The sub-bundle D defines a differential system on $M$, and, if it is integrable, the two-dimensional integral submanifolds define a codimension-two foliation of $M$ that represents the instantaneous isophase surfaces of the electromagnetic wave motion defined by $F$. The integrability criterion is given by Frobenius:

(8.5) $\quad \alpha \wedge d\alpha = \beta \wedge d\beta = 0.$

An important subtlety to consider is that if $\alpha$ and $\beta$ are globally linearly independent then they also define a global 2-frame field on $M$. Consequently, if this is true then $M$ must be topologically inclined to admit such a field, i.e., it must have a degree of parallelizability of two. A necessary condition for this is that the top two Stiefel-Whitney classes of $T(M)$, $w_3$ and $w_4$, must vanish. Otherwise, $\alpha$ and $\beta$ can only be defined on the complement of some singularity set.

Now that we have a codimension-two foliation the next issue to address is whether we can also define the complementary foliations whose intersections are the leaves of the latter foliation. However, the author admits that the answer to this question is not clear as of yet and defers that effort to further researches in order to prevent the present discussion from overstepping its scope.

**9. Discussion.** In conclusion, we shall summarize the most promising directions for further research into the topological nature of electromagnetism.

*a*) The representation of $F$ and $J$ by more fundamental geometrical or topological objects: It is not unreasonable to expect that the introduction of an $\mathfrak{sl}(4)$ connection on the $SL(4)$-principal bundle on $M$ defined by a choice of $\mathcal{V}$ is unavoidable. In conventional electrodynamics when we use a potential 1-form $A$ for $F$ as the fundamental object the usual considerations of wanting the field theory to be invariant under local $U(1)$ gauge changes dictate that $A$ must define a $U(1)$ connection 1-form. However, the cautious reader will note that the group that plays the fundamental role in topological electromagnetism seems to be $SL(4)$, not $SO(2)$. Resolving this source of confusion seems to be key to further progress.

*b*) Reduction from $SL(M)$ to $SO(3,1)(M)$: This is an essential step from the standpoint of introducing gravitation into the model, or rather, deducing gravitation from it.  We have seen that such a reduction can follow from starting with a more general scalar product on 2-vector fields, but that only transfers the burden of significance to the choice of a constitutive law.  One then must investigate the physical and mathematical bases for this choice in finer detail.

*c*) Role of the intersection form in constitutive laws:  Although the constitutive law that we just mentioned certainly gives a simple and direct route for effecting the reduction from *SL*(*M*) to *SO*(3,1)(*M*), it nevertheless lacks an immediate topological construction.  This either suggests that one cannot find a completely topological formulation for electromagnetism and must eventually resort to geometrical axioms or that we need to look for a more topological basis for the constitutive law.  Since the intersection form seems to play a role vis-à-vis $H^2(M;\mathbb{R})$ that is analogous to the role played by the constitutive law in the context of $\Lambda^2(M)$, this seems to be a promising direction to investigate.

The author wishes to thank F.W. Hehl of the University of Cologne for directing his attention to the prior attempts to represent electromagnetism in a non-metric way, also to acknowledge the University of Wisconsin at River Falls for providing a congenial and supportive research environment.